\documentclass[twocolumn]{aastex63}

\usepackage[T1]{fontenc}
\usepackage{bm}        
\usepackage{amssymb, amsmath}   
\usepackage{cancel}
\usepackage{color}

\DeclareMathAlphabet{\mathsfit}{\encodingdefault}{\sfdefault}{m}{sl}
\SetMathAlphabet{\mathsfit}{bold}{\encodingdefault}{\sfdefault}{bx}{sl}
\newcommand{\vect}[1]{\bm{#1}}

\shorttitle{Blazar Synchrotron from Turbulence}
\shortauthors{Zhang et al.}

\begin{document}

\title{First-Principle-Integrated Study of Blazar Synchrotron Radiation and Polarization Signatures from Magnetic Turbulence}

\correspondingauthor{Haocheng Zhang}
\email{haocheng.zhang@nasa.gov}

\author[0000-0001-9826-1759]{Haocheng Zhang}
\affiliation{NASA Postdoctoral Program Fellow}
\affiliation{NASA Goddard Space Flight Center\\
Greenbelt, MD 20771, USA}

\author[0000-0001-7396-3332]{Alan P.
\ Marscher}
\affiliation{Institute for Astrophysical Research\\
Boston University\\
Boston, MA 02215, USA}

\author[0000-0003-4315-3755]{Fan Guo}
\affiliation{Theoretical Division \\
Los Alamos National Lab \\
Los Alamos, NM 87545, USA}

\author[0000-0003-1503-2446]{Dimitrios Giannios}
\affiliation{Department of Physics and Astronomy \\
Purdue University \\
West Lafayette, IN 47907, USA}

\author[0000-0001-5278-8029]{Xiaocan Li}
\affiliation{Dartmouth College\\
Hanover, NH 03750, USA}

\author[0000-0002-6548-5622]{Michela Negro}
\affiliation{University of Maryland, Baltimore County\\
Baltimore, MD 21250, USA}
\affiliation{NASA Goddard Space Flight Center\\
Greenbelt, MD 20771, USA}
\affiliation{Center for Research and Exploration in Space Science and Technology, NASA/GSFC\\
Greenbelt, MD 20771, USA}

\begin{abstract}
Blazar emission is dominated by nonthermal radiation processes that are highly variable across the entire electromagnetic spectrum. Turbulence, which can be a major source of nonthermal particle acceleration, can widely exist in the blazar emission region. The Turbulent Extreme Multi-Zone (TEMZ) model has been used to describe turbulent radiation signatures. Recent particle-in-cell (PIC) simulations have also revealed the stochastic nature of the turbulent emission region and particle acceleration therein. However, radiation signatures have not been systematically studied via first-principle-integrated simulations. In this paper, we perform combined PIC and polarized radiative transfer simulations to study synchrotron emission from magnetic turbulence in the blazar emission region. We find that the multi-wavelength flux and polarization are generally characterized by stochastic patterns. Specifically, the variability time scale and average polarization degree (PD) are governed by the correlation length of the turbulence. Interestingly, magnetic turbulence can result in polarization angle (PA) swings with arbitrary amplitudes and duration, in either direction, that are not associated with changes in flux or PD. Surprisingly, these swings, which are stochastic in nature, can appear either bumpy or smooth, although large amplitude swings ($>180^{\circ}$) are very rare, as expected. Our radiation and polarization signatures from first-principle-integrated simulations are consistent with the TEMZ model, except that in the latter there is a weak correlation, with zero lag, between flux and degree of polarization.
\end{abstract}



\section{Introduction} \label{sec:intro}

Blazars are the most prominent extragalactic $\gamma$-ray sources \citep{Abdollahi2020}. They consist of rapidly accreting supermassive black hole systems that produce relativistic plasma jets pointing very close to our line of sight. Their emission is nonthermal-dominated, characterized by two spectral components \citep[see e.g.,][for a recent review]{Boettcher2019}. The low-energy component extends from radio to optical, and in some cases up to soft X-rays. It is dominated by synchrotron emission from ultra-relativistic electrons in a partially ordered magnetic field. This is evident from the observed radio to optical polarization signatures \citep[e.g.,][]{Marscher2008,Blinov2018}. The high-energy spectral component extends from X-rays up to TeV $\gamma$-rays, which may originate from either Compton scattering by the same electrons that emit the low-energy synchrotron component \citep[referred to as the leptonic scenario; see, e.g.,][]{Dermer1992,Sikora1994,Marscher1985} or synchrotron emission from ultra-relativistic protons and hadronic cascades \citep[referred to as the hadronic scenario; see, e.g.,][]{Mannheim1993,Mucke2003}. While the two scenarios cannot be distinguished by spectra alone \citep{Boettcher2013}, the recent detection of a very high energy neutrino event coinciding with the flaring blazar TXS~0506+056 provides the first strong evidence for hadronic processes in blazars \citep{IceCube2018}. These phenomena imply that blazar jets are among the most powerful particle accelerators in the Universe, and are strong candidates for the acceleration sites of ultra-high-energy cosmic rays. The currently operating {\it Imaging X-ray Polarimetry Explorer} ({\it IXPE}\footnote{\url{https://www.nasa.gov/mission_pages/ixpe/index.html}}) and future X-ray and MeV $\gamma$-ray telescopes with polarimetry capability, such as {\it eXTP}, {\it COSI}\footnote{\url{https://cosi.ssl.berkeley.edu/}}, and {\it AMEGO-X} \citep{Weisskopf2016,ZhangSN2016,Tomsick2022,Caputo2022}, can probe cosmic-ray acceleration and neutrino production via high-energy polarization signatures \citep{Zhang2013,Zhang2017b,Paliya2018,Zhang2019}.

Blazar emission is highly variable at all wavelengths. GeV and TeV-range $\gamma$-rays can flare within a few minutes in some extreme events \citep{Albert2007,Aharonian2007,Ackermann2016}. The fast variability implies that nonthermal particle acceleration occurs locally in the flaring region, often referred to as the blazar zone. Three physical processes are considered likely causes of particle acceleration: shocks, magnetic reconnection, and turbulence. Previous theoretical studies of emission from shocks have produced multi-wavelength spectra and variability consistent with blazar observations \citep{Joshi2007,Chen2014,Zhang2016,Spada2001}. This is due to strong acceleration at shocks via first-order Fermi acceleration\citep{Achterberg2001,Spitkovsky2008,Summerlin2012}. However, the acceleration is efficient only if the blazar zone is weakly magnetized. In a strongly magnetized environment, magnetic reconnection between oppositely directed magnetic field lines can efficiently accelerate particles \citep[see][for a recent review]{Guo2020}. This can happen, for example, in a kink-unstable jet or a striped jet \citep{Giannios2006,BarniolDuran2017,Giannios2019,Zhang2021}. Recent numerical studies have also illustrated the potential of explaining blazar flaring activities with reconnection \citep{Guo2014,Guo2015,Guo2021,Sironi2014,Petropoulou2019,Werner2021}. In particular, reconnection can result in rapid variability in both flux and polarization, owing to anisotropic and/or inhomogeneous distributions of nonthermal particles in the reconnection region \citep{Giannios2009,Zhang2022}. Such fast variability patterns typically occur during strong blazar flares, but are less common during weak flares or quiescent states. Stochastic acceleration has long been considered as a key mechanism involved in both flaring and quiescent states of blazars \citep[e.g.,][]{Marscher2014}. Magnetic turbulence can accelerate particles near the black hole and in the jet both from second-order Fermi acceleration and relatively minor but numerous magnetic reconnection events \citep[e.g.,][]{Dermer1996,Yan2012,Marscher2022}. Recent PIC simulations provide more insight on this process in a magnetized environment \citep{Zhdankin2017,Zhdankin2019,Comisso2018,Comisso2019}. Relativistic turbulence can accelerate nonthermal particles on a relatively fast time scale comparable to the light crossing time. While shocks, reconnection, and turbulence can all accelerate high-energy particles and explain typical blazar emission patterns, they require very different physical conditions that are essential toward understanding relativistic jet dynamics and evolution. Interestingly, the three physical processes involve distinct magnetic field evolution, which can be studied with polarization signatures.

The observed optical polarization is highly variable in many blazars \citep{Smith2009,Marscher2010,Covino2015,Marscher2021}.
The PD typically ranges from near zero to tens of percent during periods of both high and low flux states. The PA varies erratically in many blazars and fluctuates by tens of degrees about the jet direction. In some blazars, the PA rotates over tens or hundreds of degrees \citep{Smith2009,Marscher2021}. Importantly, some blazar flares are simultaneous with large optical PA swings, indicating significant magnetic field evolution correlated with particle acceleration \citep{Blinov2015,Morozova2014,Dammando2011,Ikejiri2011}. These rotations can happen in both directions, and in some cases go far beyond $180^{\circ}$, which is unlikely to be caused by purely geometric effects \citep{Chandra2015,Marscher2010}. Both deterministic and stochastic models have been put forward to explain such rotations \citep[see e.g.,][for a recent review]{Zhang2019b}. Generally speaking, shocks and reconnection can both lead to PA swings simultaneous with flares, but the shock scenario predicts a maximum swing of $\sim 180^{\circ}$ for a single shock, and the PD can either rise or drop during the swing \citep{Chen2014,Zhang2016}. In the cases with no PA swings, a shock usually increases the PD, and cannot give erratic patterns without the introduction of turbulence \citep{Laing1980,Tavecchio2020}. On the other hand, reconnection predicts fast angle swings, in either direction, that can extend beyond $180^{\circ}$ \citep{Zhang2020,Zhang2015}. They are associated with plasmoid mergers in the reconnection region, which also give rise to multi-wavelength flares \citep{Zhang2022}. While the above signatures associated with reconnection seem to match well extreme blazar flares with optical PA swings, such events are observed to occur a small fraction of the time. Therefore, the question remains as to what physical mechanisms drive more typical blazar activities.

It is possible to combine the turbulence and shock scenarios, as in the TEMZ model \citep{Marscher2014}, in which turbulent cells of plasma cross a shock front. A key feature of the model is that the electrons are accelerated to the highest energies most efficiently in locations where the magnetic field $\bf B$ is nearly parallel to the shock normal, the so-called ``subluminal" regime within which particles can pass back and forth across the shock front multiple times \citep[e.g.,][]{Summerlin2012}). This limits the effective volume of emission at the highest frequencies, emitted by the highest-energy particles. This effect, combined with the spatial variations of physical parameters because of the turbulence, causes the flux and polarization to fluctuate more strongly at higher frequencies. Turbulence can also be generated in a magnetized blazar environment due to magnetic instabilities, where particles are accelerated by dissipating magnetic energy. Similar to the TEMZ model, the highest energy particles can only occupy a small region near the magnetic energy dissipation sites, thus higher-frequency emission can appear more variable than at lower frequencies. 

This paper aims to study the synchrotron radiation and polarization signatures from turbulence in the blazar zone environment. We use coupled PIC and polarized radiative transfer simulations to study the plasma dynamics, particle acceleration, radiation, and feedback from first principles. Our goal is to both identify general radiation and polarization patterns and explore potentially unique signatures from turbulence, which can be distinctive from shock and reconnection scenarios. Additionally, we emphasize which physical quantities determine the temporal evolution of radiation and polarization signatures. While our combined PIC and radiative transfer simulations mainly consider turbulence in a magnetized blazar zone, we will compare our results with the TEMZ model to explore any differences that might exist between magnetic- and kinetic-driven turbulence. Section \ref{sec:setup} describes our simulation setup, \S\ref{sec:results} presents general observable signatures, \S\ref{sec:PAswings} examines two cases with PA swings, \S\ref{sec:correlationlength} focuses on the effects of turbulence correlation length, \S\ref{sec:parameterstudy} performs additional parameter studies, \S\ref{sec:temz} compares our results with a simulation of the TEMZ model, \S\ref{sec:implication} discusses implications for observations, and we summarize our results in \S\ref{sec:discussion}.

\section{Simulation Setup \label{sec:setup}}

We assume that the turbulence exists in a substantially magnetized blazar zone environment. We will show in \S\ref{sec:temz} that the general radiation and polarization signatures from kinetic-driven turbulence, as simulated via the TEMZ model, appear quite similar to the magnetic-driven turbulence presented in the following sections. The blazar zone has a bulk Lorentz factor $\Gamma=10$. The PIC simulations are performed in the co-moving frame of the bulk flow. The simulations start from a relativistic thermal bath with no initial nonthermal particles. Turbulence is triggered by injecting an ensemble of magnetic fluctuations at scales close to the domain size. We consider various initial parameters to study the dependence of radiation and polarization signatures on the physical conditions of turbulence. Both our PIC and TEMZ simulations include radiative cooling from both synchrotron and Compton scattering, which are the dominating particle cooling mechanisms under the leptonic blazar model. Since we only focus on the synchrotron emission in this paper, we do not explore the ratio between synchrotron and Compton scattering cooling, but make the two comparable for all our simulations. In the following subsection, we describe in detail our PIC and radiative transfer setups.

\subsection{PIC Setup}

We carry out 2D relativistic turbulence simulations using the \texttt{VPIC} particle-in-cell code \citep{Bowers2008}, which solves the relativistic Vlasov-Maxwell system of equations. Similar to earlier studies \citep{Comisso2018,Comisso2019}, the simulations start from a uniform mean magnetic field $B_0\hat{\vect{z}}$ and a spectrum of magnetic fluctuations $\delta\vect{B}$ in the $x$--$y$ plane, with $\delta B^2_\text{rms0}\equiv\left<\delta B^2\right>_{t=0}=B_0^2$ and $\delta\vect{B}(\vect{r})=\sum_{\vect{k}}\delta B(\vect{k})\hat{\vect{\xi}}(\vect{k})\exp[i(\vect{k}\cdot\vect{r}+\phi_{\vect{k}})]$, where $\delta B(\vect{k})$ is the amplitude of each wave mode, $\hat{\vect{\xi}}(\vect{k})\equiv i\vect{k}\times\vect{B}_0/|\vect{k}\times\vect{B}_0|$ is the polarization unit vector, and $\phi_{\vect{k}}$ is the wave phase. The wave vector $\vect{k}=(k_x, k_y)$, where $k_x=2\pi m/L_x$ and $k_y=2\pi n/L_y$ for a domain size of $L_x\times L_y$ and $m\in\{-N_x,\cdots-1,1\cdots N_x\}$ and $n\in\{-N_y,\cdots-1,1\cdots N_y\}$. $N_x$ and $N_y$ are the number of modes along each direction. In the rest of the discussion, we adopt $L_x=L_y=2L$ and $N_x=N_y=N$. The simulation domain is $[-L<x< L, -L<y<L]$, with the default choice of $L=16000d_{e0}$ and $N=8$, where the electron inertial length $d_{e0}=c/\omega_{pe0}$, where $\omega_{pe0}\equiv\sqrt{4\pi n_0e^2}/m_e$ is the non-relativistic electron plasma frequency. The wave phases are assumed to be random within 0 and $2\pi$. To ensure that $\delta\vect{B}$ is real, we assume $\delta B(-\vect{k})=\delta B(\vect{k})$ and $\phi_{-\vect{k}}=-\phi_{\vect{k}}$. If each wave mode carries the same power \citep[equal amplitude per mode, similar to][]{Comisso2019}, $\delta B(\vect{k})=\delta B_\text{rms0}/2N$. Although the initial wave phases do not affect the physical evolution of the turbulence, they can lead to different radiation signatures, as found in our simulations.

We perform the simulations in a proton-electron plasma with a physical mass ratio $m_i/m_e=1836$. The initial particle distributions are Maxwell–J\"uttner distributions with uniform density $n_0$, dimensionless temperature $\theta_e=kT_e/m_ec^2=50$, and $T_i=T_e$, so that the electrons are relativistic, while the protons are non-relativistic. We set the cold electron magnetization parameter $\sigma_{e0}\equiv B_0^2/(4\pi n_0m_ec^2)$ (default $\sigma_{e0}=4\times 10^4$, or total $\sigma\sim 22$), defined using the mean magnetic field. As a result, $\omega_{pe0}/\Omega_{ce0}=1/200$, where $\Omega_{ce0}\equiv eB/m_ec$ is the non-relativistic electron gyrofrequency. The simulation box is resolved using grids with $n_x\times n_y=8192\times8192$. The resulting grid sizes $\Delta x=\Delta y\approx3.9d_{e0}\approx0.55d_e$, where $d_e\equiv d_{e0}/\sqrt{\theta_e}$ is the electron inertial length including the relativistic correction. We use 100 particles per species per cell. For both fields and particles, we employ periodic boundaries along the $x$ and $y$ directions. We implement a radiation reaction force to mimic the cooling effect in blazars, which can be considered as a continuous frictional force for relativistic electrons \citep[non-relativistic terms are ignored; see][]{Cerutti2012,Cerutti2013},
\begin{align*}
  \vect{g}
  = -\frac{2}{3}r_e^2\gamma\left[\left(\vect{E}+
  \frac{\vect{u}\times\vect{B}}{\gamma}\right)^2 -
  \left(\frac{\vect{u}\cdot\vect{E}}{\gamma}\right)^2\right]\vect{u}-\frac{4}{3}\sigma_T\gamma \mathcal{U}_{\star}\vect{u},
\end{align*}
where $\vect{u}=\gamma \vect{v}/c$ is the four-velocity, $r_e=e^2/m_ec^2$ is the classical radius of the electron, and $\mathcal{U}_{\star}$ is the photon energy density. We assume that the Compton cooling is dominated by external photons, whose energy density is comparable to the upstream magnetic energy density. Given the fact that the typical blazar cooling parameters have trivial effects on PIC scales, we compensate the cooling force by multiplying by a factor such that the cooling break of the particle spectrum occurs at $\gamma_c\sim 10^4$. With this setup, we estimate that the so-called radiation-reaction (burn-off) limit, where the cooling becomes comparable with the Lorentz force, is at $\gamma_{rad}\sim 6 \times 10^4$ \citep{Uzdensky2011}.

We present in the following two sections three cases with different initial phase realizations, labeled as Case 1, 2, and 3. Most of our simulations behave similarly to Case 1, where radiation and polarization signatures appear stochastic. But Cases 2 and 3 exhibit significant polarization variations. \S\ref{sec:correlationlength} and \S\ref{sec:parameterstudy} present additional runs with the same initial phase realization as for Case 2, but with different physical parameters (different radiation transfer resolutions, simulation box sizes, and number of modes in \S\ref{sec:correlationlength}, different magnetization factor $\sigma$ and cooling factor in \S\ref{sec:parameterstudy}). Their differences are listed in Table \ref{tab:pic}.

\begin{table}
\centering
\begin{tabular}{|l|c|c|c|c|c|}
\hline
Run \#  &  $\sigma_{e0}$     &  $f_{cool}$   &  Size ($d_{e0}$)      &  Phase          &  $N$  \\ \hline
Case1   &  $4\times 10^4$    &  200          &  $32000^2$            &  0              &  8    \\ \hline
Case2   &  $4\times 10^4$    &  200          &  $32000^2$            &  16384          &  8    \\ \hline
Case3   &  $4\times 10^4$    &  200          &  $32000^2$            &  81920          &  8    \\ \hline
Case2a  &  $4\times 10^4$    &  200          &  $24000^2$            &  16384          &  8    \\ \hline
Case2b  &  $4\times 10^4$    &  200          &  $16000^2$            &  16384          &  8    \\ \hline
Case2c  &  $4\times 10^4$    &  200          &  $12000^2$            &  16384          &  8    \\ \hline
Case2d  &  $4\times 10^4$    &  200          &  $8000^2$             &  16384          &  8    \\ \hline
Case2e  &  $4\times 10^4$    &  200          &  $32000^2$            &  16384          &  6    \\ \hline
Case2f  &  $4\times 10^4$    &  200          &  $32000^2$            &  16384          &  12   \\ \hline
Case2g  &  $1.6\times 10^5$  &  200          &  $32000^2$            &  16384          &  8    \\ \hline
Case2h  &  $4\times 10^4$    &  400          &  $32000^2$            &  16384          &  8    \\ \hline
\end{tabular}
\caption{List of parameters varied in our PIC simulations: $\sigma_{e0}$ is the electron magnetization factor, while $f_{cool}$ is a cooling factor in the code, which adjusts the radiation reaction force. A larger $f_{cool}$ means weaker cooling. Size is the 2D simulation domain size in units of the electron inertial length. Phase is the code value that is used to set the initial phase of the perturbation. $N$ is the number of modes in the perturbation. Cases 1, 2, and 3 are shown in \S\ref{sec:results}, which are only different by their initial phases of the perturbation, set by 0, 1, and 5 times of the CPU numbers, respectively. Cases 2a-d have the same parameters as Case 2 but with different box sizes, and Cases 2e-f have different numbers of modes. These runs are described in \S\ref{sec:correlationlength}. Case 2g has a higher initial $\sigma$ and Case 2h has slower cooling, as discussed in \S\ref{sec:parameterstudy}.}
\label{tab:pic}
\end{table}

\subsection{Radiative Transfer Setup}

The magnetic field and particle evolution derived from PIC simulations are post-processed with the \texttt{3DPol} code developed by \citet{Zhang2014}. We fix our line of sight along the $z$-axis in the comoving frame of the simulation domain, perpendicular to the $x$--$y$ plane where the turbulence occurs. We assume that the simulation domain is moving up along the $+y$ direction with a bulk Lorentz factor $\Gamma=10$, so that the Doppler factor $\delta\equiv\Gamma=10$ in our setup. We make this choice because typical blazar observations and spectral fitting models suggest a Doppler factor in the range of ten to a few tens \citep[e.g.,][]{Boettcher2019}. The initial magnetic field strength is normalized to $0.1~\rm{G}$, which is a typical value for the leptonic scenario \citep[e.g.,][]{Boettcher2013}. To obtain adequate statistics of particle spectra for the radiative transfer simulation and save computational resources, we sum electrons in $32\times 32$ PIC cells into one radiative transfer cell. We reduce the magnetic field by averaging the magnetic field within these PIC cells, so that any disorder of the magnetic field components on scales smaller than the radiative transfer cell are ignored. Although this treatment may seem oversimplified, we find in \S\ref{sec:correlationlength} that the time evolution of radiation and polarization signatures is mostly similar as long as the turbulence correlation length is well resolved. We bin the particle kinetic energy $(\gamma_e-1)m_ec^2$ into 100 steps between $10^{-4}m_ec^2$ and $10^6m_ec^2$. We output the above information every $\sim 0.0078\tau_{lc}$ to obtain adequate temporal resolution. Under the default resolution, the \texttt{3DPol} code has a resolution of $256\times 256$. It then calculates the Stokes parameters at every time step in each radiative transfer cell, and ray-traces to the plane of the sky. In this way we can obtain the time-, space-, frequency-, and polarization-dependent synchrotron emission, allowing us to analyze the total radiation and polarization signatures, as well as the spatially resolved emission maps.

\section{General Synchrotron Radiation Signatures from Turbulence \label{sec:results}}

In this section, we study the synchrotron radiation and polarization signatures from our coupled PIC and polarized radiative transfer simulations. Our default simulation setup, Case 1, represents the majority of our simulation runs, with rather ``featureless'' temporal evolution of radiation and polarization signatures. We find that turbulence can co-accelerate nonthermal electrons and protons, but only electrons suffer from significant cooling, which leads to a cooling break at high energies. The multi-wavelength radiation and polarization signatures are generally similar, although the higher-energy band appears slightly more variable. Nonthermal particles concentrate near the edge of magnetic islands in turbulence, where the magnetic field is more ordered. Nonetheless, due to the overall disordered magnetic field morphology, the polarization degree in all wavelengths remains at a low level throughout the simulation.

\subsection{Spectral Properties}

\begin{figure}
\centering
\includegraphics[width=\linewidth]{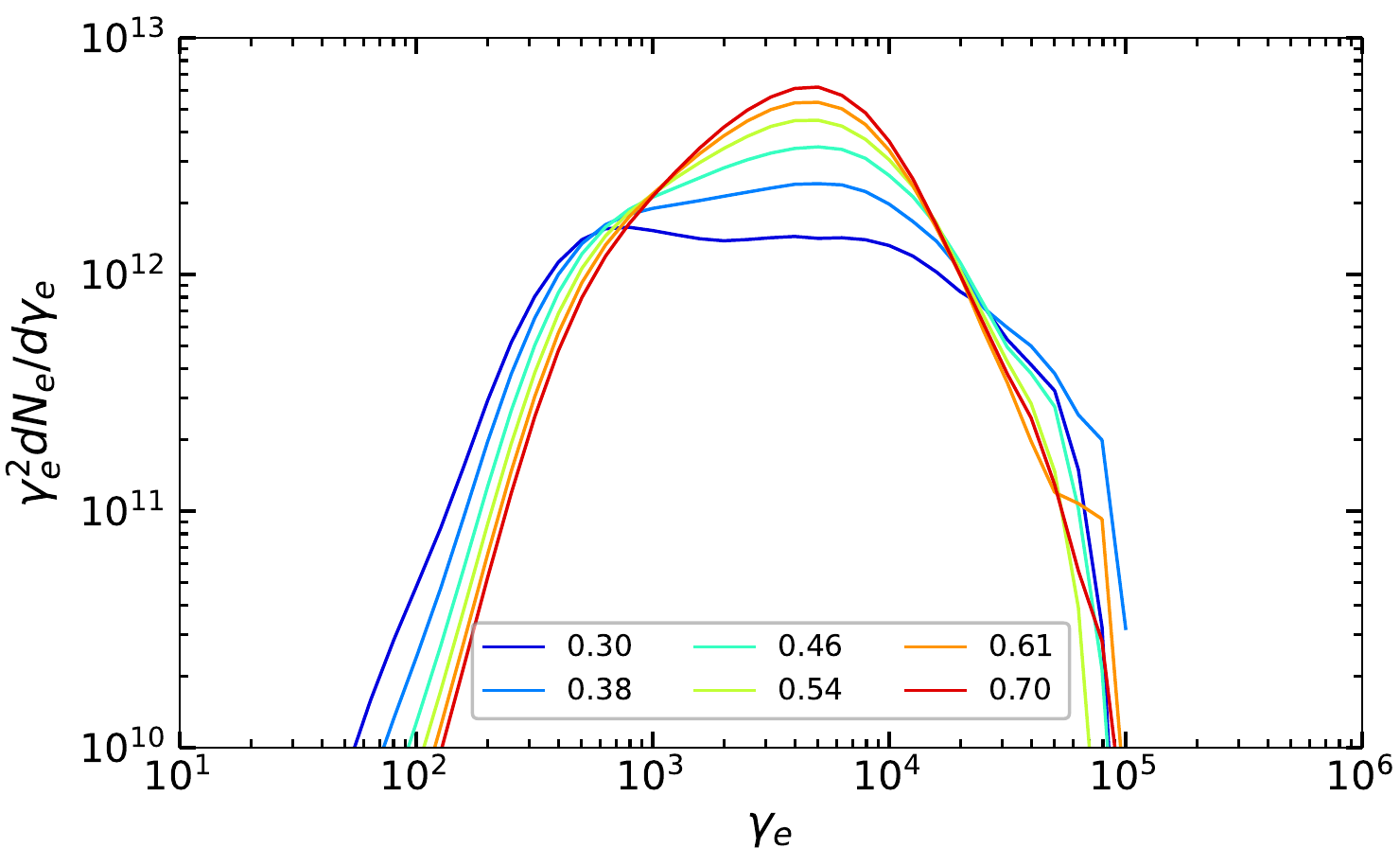}
\includegraphics[width=\linewidth]{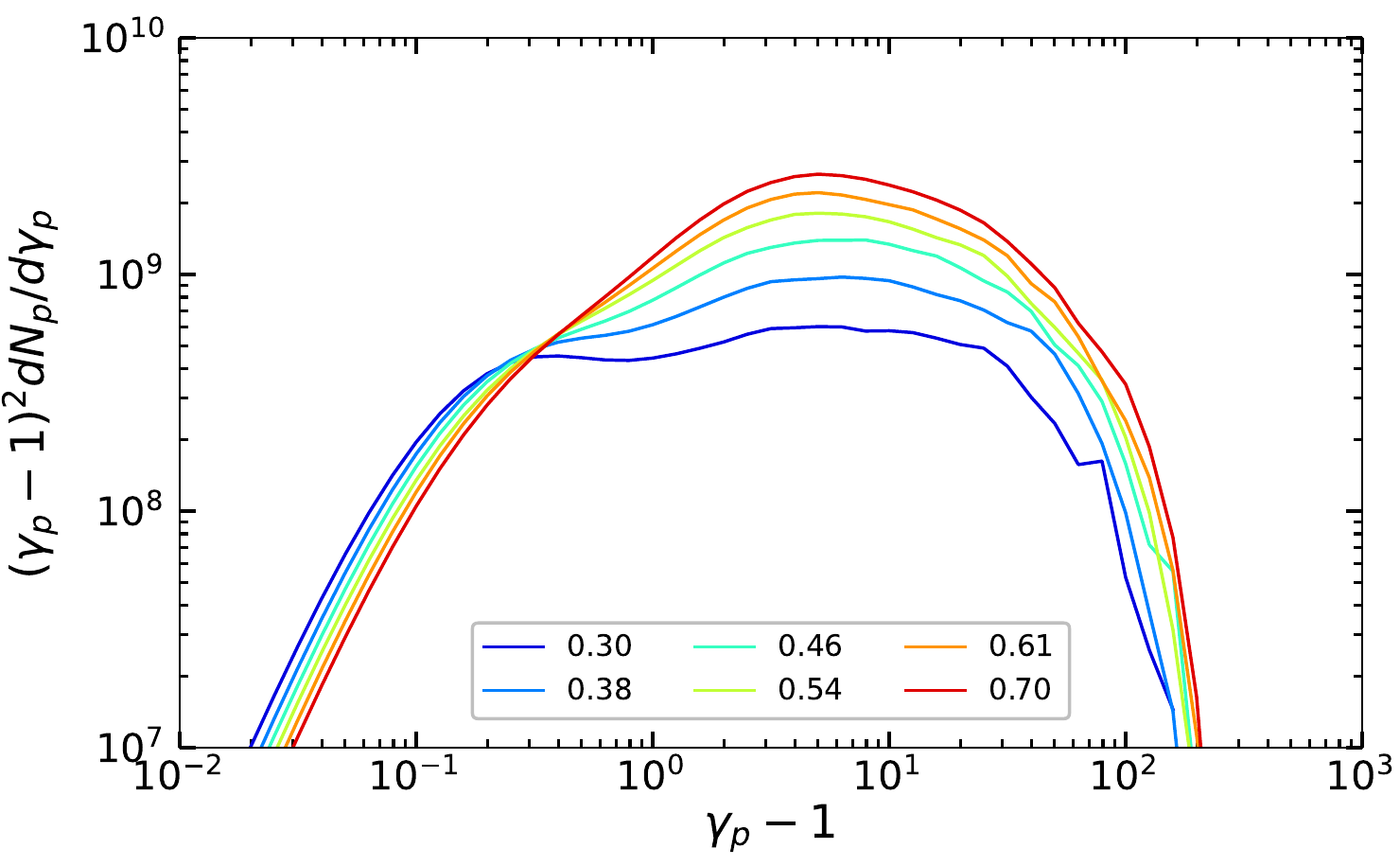}
\caption{From top to bottom: electron and proton energy distributions. Equally spaced snapshots are selected during the middle of the simulation when the turbulence has fully developed but not yet saturated. All results are presented in the co-moving frame of the simulation domain.}
\label{fig:Case1particle}
\end{figure}

\begin{figure}
\centering
\includegraphics[width=\linewidth]{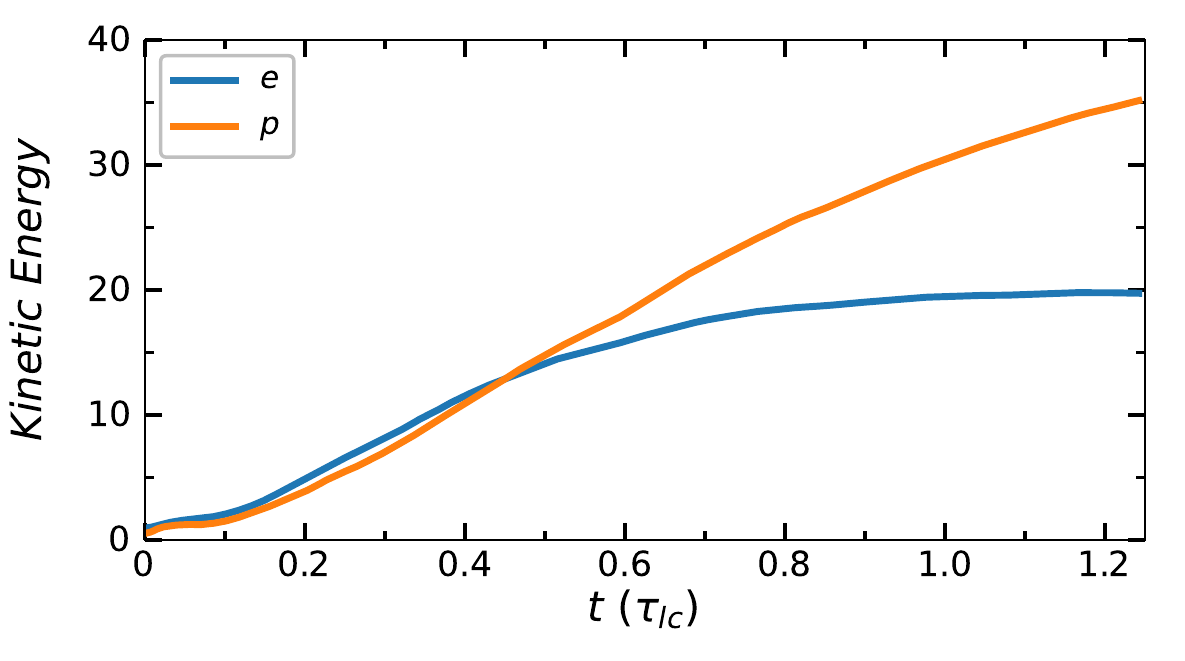}
\caption{Temporal evolution of the kinetic energy in electrons and protons. Both are presented in the co-moving frame of the simulation domain.}
\label{fig:Case1ke}
\end{figure}

Figure \ref{fig:Case1particle} shows snapshots of the electron and proton energy spectra. Turbulence can co-accelerate nonthermal electrons and protons by dissipating the magnetic energy in the simulation domain. At relatively early stages of turbulence evolution ($t\lesssim 0.4\tau_{lc}$ in Figure \ref{fig:Case1particle}), both species exhibit a power-law index of $p\sim 2.0$. The kinetic energies contained in the two species are nearly identical, as shown in Figure \ref{fig:Case1ke}. Turbulence continues to heat and accelerate protons, thus the low-energy end of the proton spectrum gradually moves to higher energy. A thermal peak also grows at $\gamma_p \sim$ a few ($t\sim 0.7\tau_{lc}$ in Figure \ref{fig:Case1particle} lower panel). There are more nonthermal protons accumulating at higher energy, and the maximum proton energy moves higher. However, the increase of kinetic energy of the electrons reaches a plateau at $t\sim 0.6\tau_{lc}$ (Figure \ref{fig:Case1ke}), indicating that a significant amount of magnetic energy has already been dissipated. This explains the slightly softer nonthermal proton spectra at later stages. The effect is more apparent for electrons: although the heating at lower energies continues, the acceleration of high-energy electrons is suppressed by radiative cooling. As a result, the spectrum softens to a power-law index of $p_e\sim 3.0$ beyond the cooling break $\gamma_c\sim 10^4$. This leads to a clear turnover in the particle spectra (upper panel in Figure \ref{fig:Case1particle}). The plateau in the electron kinetic energy after $t\sim 0.6\tau_{lc}$ suggests that the acceleration and cooling have reached quasi-equilibrium. This is reflected in the spectral evolution, where the nonthermal regions of the spectra are nearly identical at $t=0.61\tau_{lc}$ and $t=0.7\tau_{lc}$. We note that, at the highest energies, the particle spectra appear to harden slightly in some snapshots ($0.38$ and $0.61$ in Figure \ref{fig:Case1particle}). This may result from some mergers between magnetic islands in the 2D turbulence, which can locally enhance the acceleration via magnetic reconnection.

\begin{figure}
\centering
\includegraphics[width=\linewidth]{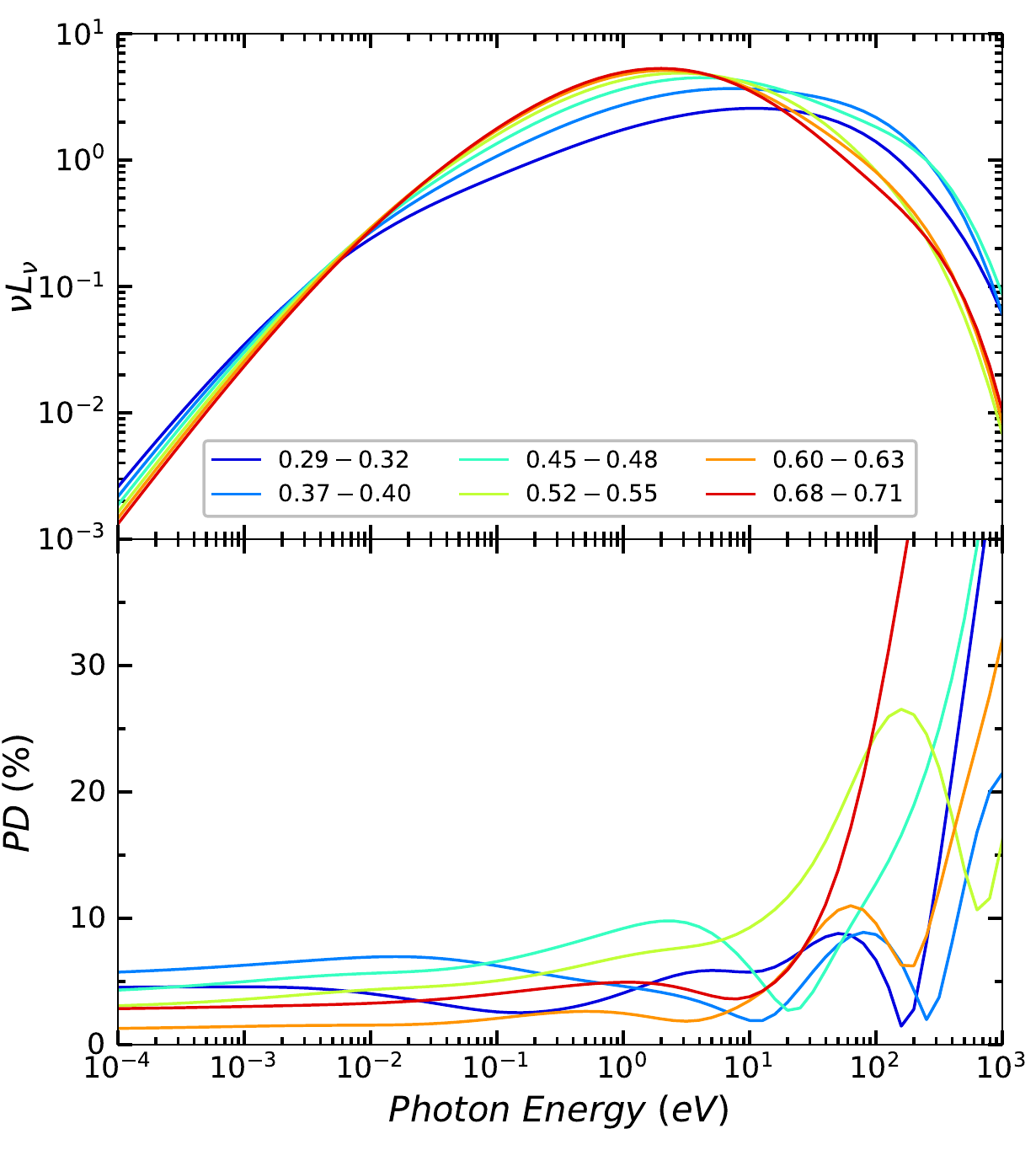}
\caption{From top to bottom: the synchrotron spectra and frequency-dependent polarization degrees (PDs). The six spectra and PDs are integrated in equal time windows centered on the selected snapshots of the electron energy distribution in Figure \ref{fig:Case1particle}. All results are present in the observer's frame. Note that the very high PDs at the spectral cutoff are due to the very soft particle spectrum and the insufficient photon statistics.}
\label{fig:Case1spec}
\end{figure}

Figure \ref{fig:Case1spec} shows snapshots of the synchrotron spectra and frequency-dependent PD. We observe that the synchrotron spectral evolution follows the electron energy distribution quite well: when the turbulence begins to develop at $t\sim 0.3\tau_{lc}$, the synchrotron spectrum peaks beyond $10~\rm{eV}$, with a rather hard spectral index $s\sim 0.5$. However, later the spectral peak is decreased by cooling, exhibiting a clear broken power-law shape with a break at $\sim 1~\rm{eV}$. Such broken power-law spectra are frequently observed in blazars. The electron heating at lower energies appears as a minor increase of the low-energy spectral cut-off frequency (upper panel in Figure \ref{fig:Case1spec}). Nonetheless, this effect may not be observed in practice, since these bands are usually dominated by the larger-scale jet emission. The PD is systematically low, implying a highly disordered magnetic field (Figure \ref{fig:Case1spec}, lower panel). The PD at $\sim 100~\rm{eV}$, which is well beyond the cooling break, is still low ($\sim 10\%$) but more variable. This suggests that the magnetic field is very disordered even on small scales, where highest-energy electrons concentrate. At even higher photon energies, since the spectra sharply cut off, there are barely enough photons to allow meaningful polarization measurements, hence their high PDs are unlikely to be observable.

\subsection{Temporal Behavior}

\begin{figure}
\centering
\includegraphics[width=\linewidth]{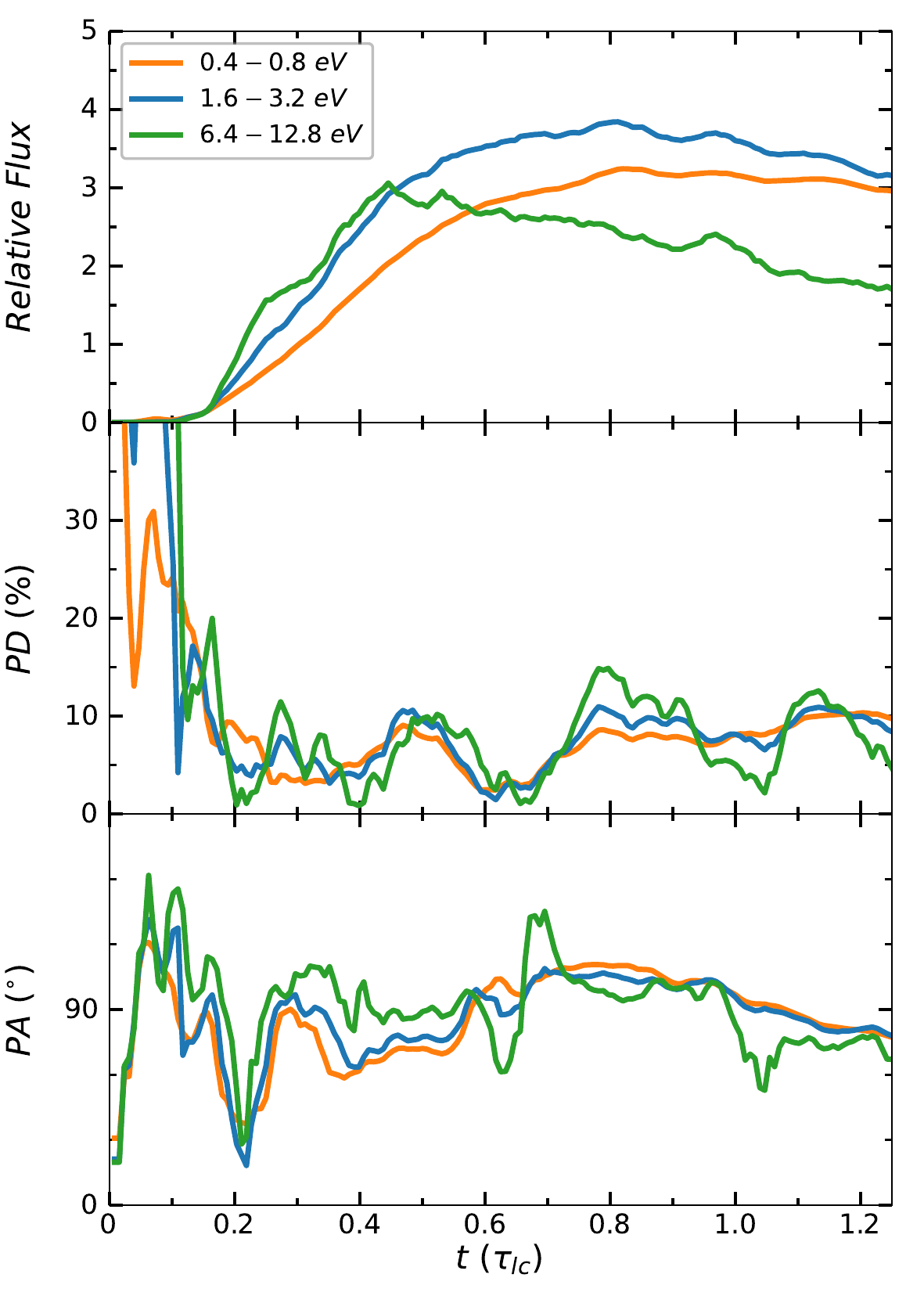}
\caption{From top to bottom: multi-wavelength light curves, time-dependent PDs, and PAs. The three bands are infrared ($0.4-0.8~\rm{eV}$), optical ($1.6-3.2~\rm{eV}$), and ultraviolet ($6.4-12.8~\rm{eV}$). Light curves are in code units, and time is in units of the light crossing time of the simulation box $\tau_{lc}$. All results are present in the observer's frame. Note that the very high PDs in the beginning of the simulation are due to insufficient photon statistics.}
\label{fig:Case1temp}
\end{figure}

Since the turbulence is triggered by a spectrum of long-wavelength magnetic fluctuations, the initial field is ordered with a very high PD, nearly 75\% (not shown). Due to insufficient photon statistics in the very beginning of the simulation, the PD remains very high for $t<0.2\tau_{lc}$.  After the turbulence develops and starts to accelerate particles into power-law distributions ($t\sim 0.2\tau_{lc}$), the optical flux gradually rises. The magnetic field morphology becomes very disordered after the initial long-wavelength perturbations break into smaller structures. Consequently, the PD quickly drops to a very low level, $\sim5$\%. Most of our simulations are similar to Case 1, where the local polarized flux in each cell of the simulation domain appears low and randomly distributed, thus the PA only fluctuates near a mean value throughout the simulation. The mean PA in Case 1, $\sim 90^{\circ}$, is just a coincidence. We will see in the following simulations that this mean value does not generally favor either $0^{\circ}$ or $90^{\circ}$. Although the radiative cooling already starts to soften the electron spectra beyond the cooling break at $t\gtrsim 0.5 \tau_{lc}$ (Figure \ref{fig:Case1spec} upper panel), the number of nonthermal electrons near the cooling break $\gamma_c\sim 10^4$ remains high. This suggests that the acceleration and cooling maintain quasi-equilibrium for a relatively long time at the spectral break. Thus after the optical flux reaches its maximum at $t\sim 0.8\tau_{lc}$, it remains at this level until $t\sim \tau_{lc}$, when the flux gradually drops due to insufficient particle acceleration. Given the rather stochastic patterns in flux and polarization, we suggest that the radiation and polarization signatures in turbulence can be approximated by low-amplitude random walks in simulation cells, similar to the TEMZ model \citep[][see \S\ref{sec:temz} for comparison]{Marscher2014}.

Although the overall signatures are similar, the ultraviolet band appears slightly more variable in flux and polarization than the other two bands in Figure \ref{fig:Case1temp}. We compare the infrared, optical, and ultraviolet bands because they represent three spectral positions in Figure \ref{fig:Case1spec}: the infrared band is below the synchrotron cooling break, the optical band is near the spectral turnover, while the ultraviolet band is above the cooling break. Their differences can be attributed to radiative cooling. Since the electrons that give rise to higher-energy synchrotron emission suffer from stronger cooling, especially true for the electrons that make ultraviolet emission, which is above the cooling break, we expect that the cooling can significantly reduce their number and the resulting synchrotron flux. Therefore, the ultraviolet band represents the turbulent regions that have the strongest particle acceleration, which likely vary more strongly than the regions that emit more in the other two bands.

\subsection{Spatial Patterns}

\begin{figure*}
\centering
\includegraphics[width=\linewidth]{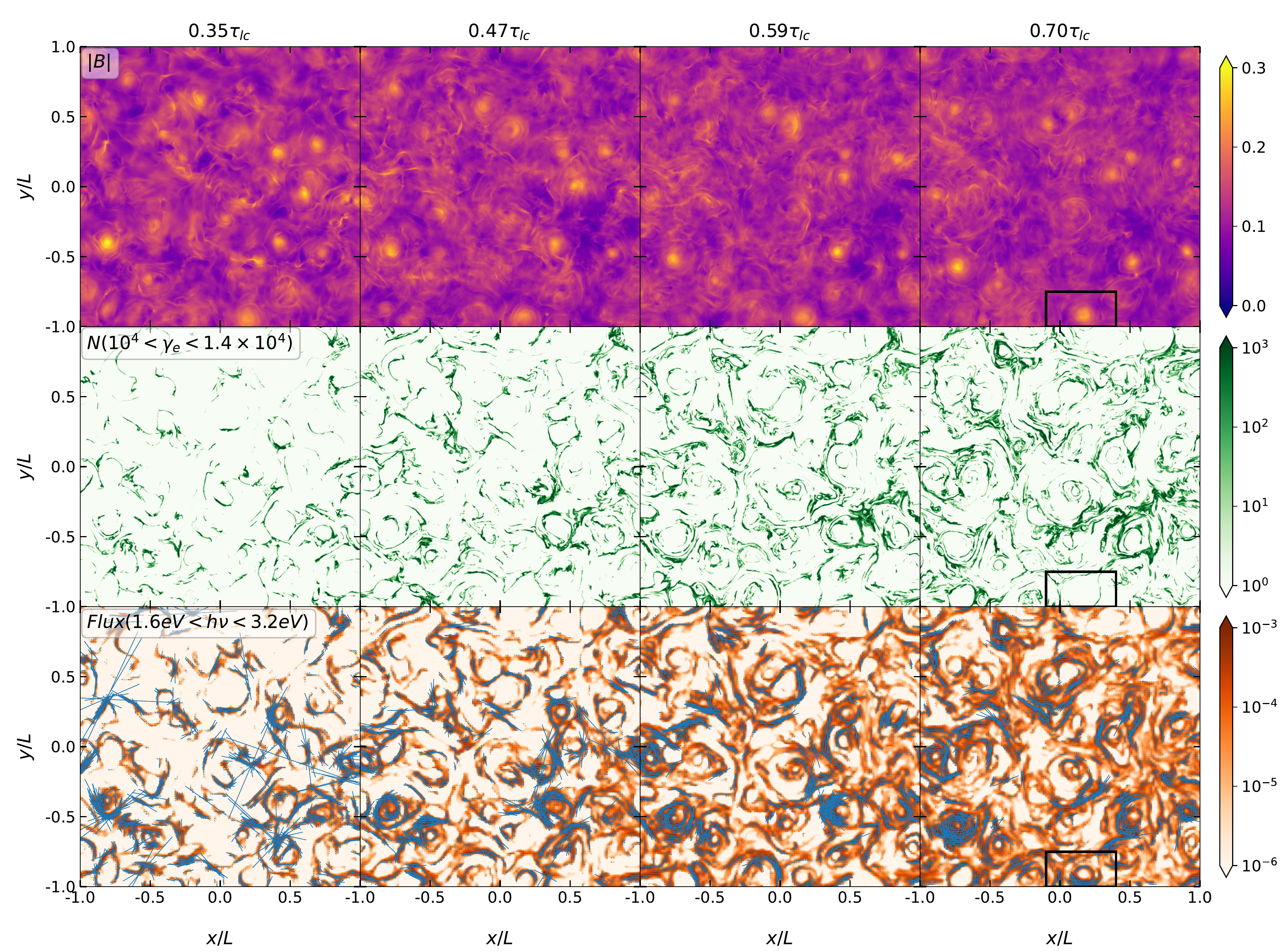}
\caption{Columns from left to right: four snapshots of the turbulence evolution. Rows from top to bottom are the magnetic field, nonthermal particle spatial distributions, and polarized intensity maps. The magnetic field strength is in Gauss, while the other two are in code units. The nonthermal particles are selected in the Lorentz factor range that corresponds to the optical band for an average magnetic field of $0.1~\rm{G}$. The length and direction of blue dashes in the last row represent the local polarized intensity and angle, respectively. One magnetic island is highlighted in the black box in the last snapshot. It is apparent that the island is at a similar position in previous snapshots. All quantities are present in the comoving frame of the emission region.}
\label{fig:Case1map}
\end{figure*}

\begin{figure*}
\centering
\includegraphics[width=\linewidth]{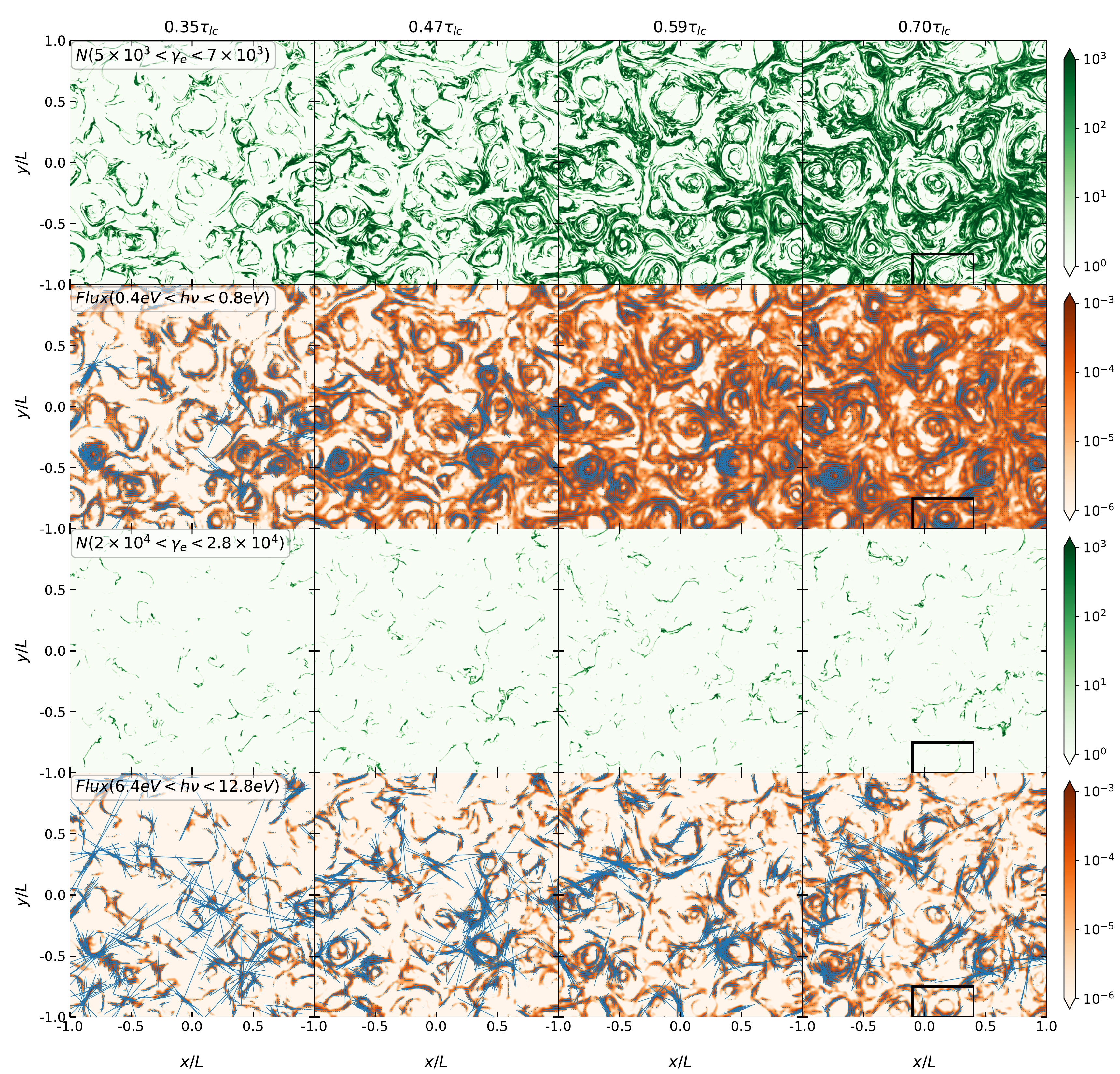}
\caption{Snapshots of the particle spatial distributions and polarized emission maps. The top two rows show the particles responsible for the infrared band and the corresponding polarized emission maps, while the bottom two rows display the ultraviolet band. The snapshots are plotted in the same way as in the bottom two rows of Figure \ref{fig:Case1map}.}
\label{fig:Case1mapmultiband}
\end{figure*}

\begin{figure}
\centering
\includegraphics[width=\linewidth]{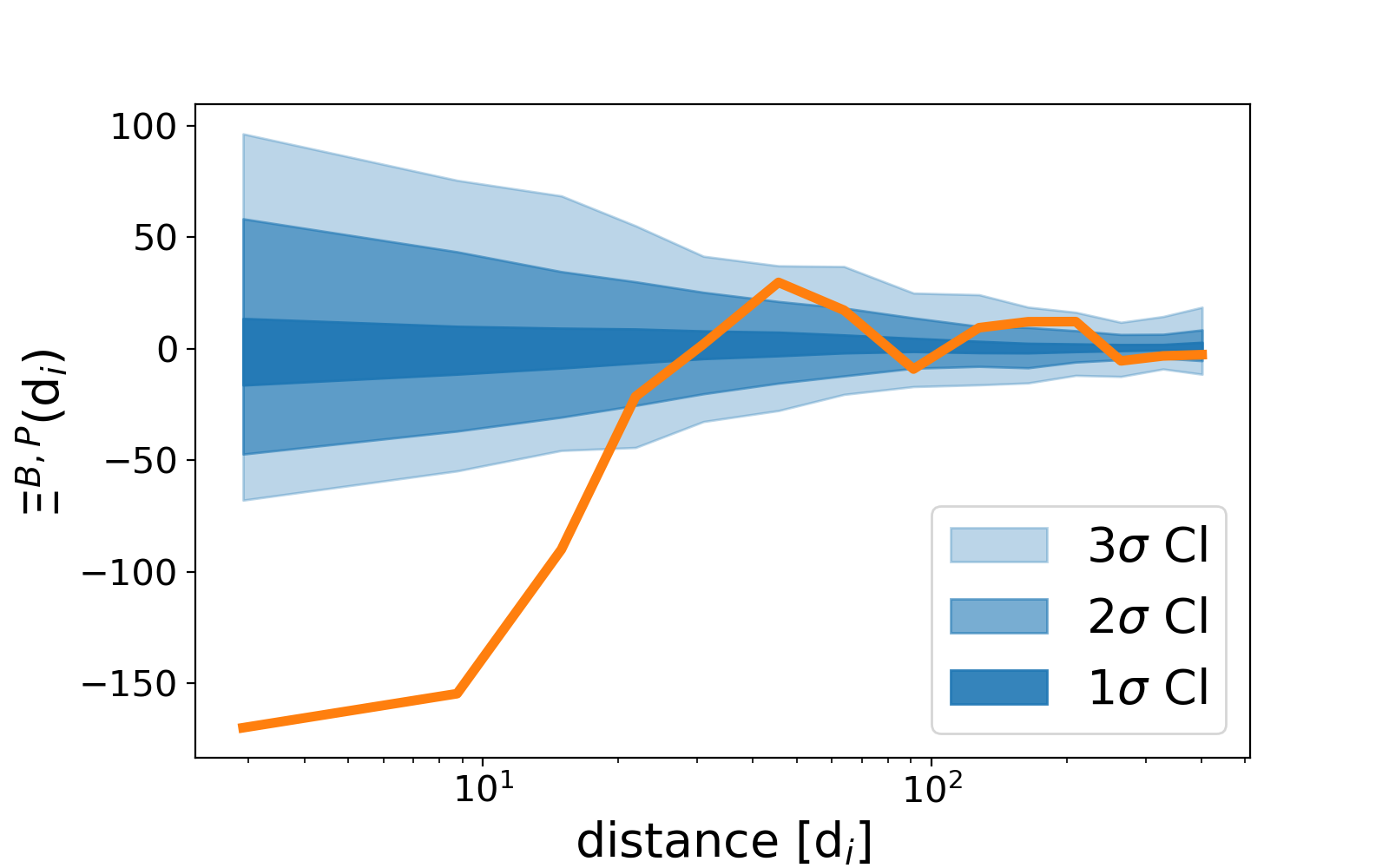}
\caption{Cross-correlation between the magnetic field and nonthermal electron density for the last snapshot in Figure \ref{fig:Case1map}. The cross-correlation function, defined in Equation~\ref{eq:ccf}, has arbitrary units. The orange lines refer to the correlation that centers on the magnetic islands. The blue bands correspond to the 68\%, 95\% and 99.7\% confidence levels for the null-hypothesis (no expected spatial correlation), which are obtained by running 2k cross-correlation function realizations centered at random points away from the centers of magnetic islands.}
\label{fig:cross-corr}
\end{figure}

After the turbulence develops, we observe some magnetic islands, which are local structures with strong magnetic field concentrated at the center (Figure \ref{fig:Case1map} first row, yellow structures are islands, and the black box in the fourth column highlights one island). By comparing the magnetic field and nonthermal electron distributions (Figure \ref{fig:Case1map} first and second rows) in these magnetic islands, it appears that electrons are concentrated on the outskirts of the islands. The anti-correlation between magnetic field strength and the energetic particles is expected, since the system is nearly in pressure balance. To quantify this connection, we use an analytical computation inspired by the 2-point cross-correlation function, typically used to measure the statistical correlation between the spatial distributions of galaxies (see, e.g., \cite{cuoco2018}). We define the cross-correlation estimator, $\Xi^{\Phi_1,\Phi_2}$, as a function of the ion skin depth ${\rm d}_i$:
\begin{equation}
    \Xi^{\Phi_1,\Phi_2}(\Delta{\rm d}_i) = \frac{1}{N_{\Delta{\rm d}_i}} \sum_{j,k}\frac{(\Phi_1^j - \left< \Phi_1\right>)}{\left< \Phi_1\right>}\frac{(\Phi_2^k - \left< \Phi_2\right>)}{\left<\Phi_2\right>},
    \label{eq:ccf}
\end{equation}
where $\Phi_{1,2}$ represents either the magnetic field $B$ or the nonthermal particle distribution. The bracketed quantities $\left< \cdot \right>$ denote the average value over the simulation domain. The quantity $\frac{(\Phi_1^j - \left< \Phi_1\right>)}{\left< \Phi_1\right>}$ assures that we are dealing with unitless fluctuation fields, which are unrelated to the absolute values of the quantities considered. The variable $\Delta{\rm d}_i$ is the distance that defines an annulus around the magnetic island (the central pixel is manually selected from the image of the magnetic field 2D realization). The factor $N_{{\rm d}_i}$ normalizes the number of pixels in each annulus considered in the computation of $\Xi$. This estimator gives the cross-correlation between $\Phi_1$ and $\Phi_2$ at the selected locations. The results are given in Figure~\ref{fig:cross-corr}. The blue shaded regions are computed by selecting $\sim$2000 sets of random points far from the magnetic islands. These ``background'' runs determine the 1, 2, and 3$\sigma$ confidence intervals within which a given value of correlation/anti-correlation occurs by chance. The orange line in Figure~\ref{fig:cross-corr} shows a clear ($\gg 99.7\%$ C.l.) anti-correlation between the magnetic field and nonthermal particle density within a distance of $\sim 20\rm{d_i}$. This clearly demonstrates that the nonthermal particle density is anti-correlated with magnetic field strength within magnetic islands within the turbulent plasma. It switches to a fairly significant correlation ($> 95\%$ C.l.) at $\sim 50\rm{d_i}$ from the center of magnetic islands. Interestingly, the $2\sigma$ correlation at $\sim 50\rm{d_i}$ occurs at approximately the smallest wavelength of the initial magnetic fluctuations in the simulation domain, $L/N= 47d_i$, which is essentially the average size of magnetic islands.

Figure \ref{fig:Case1map} (third row) shows that the optical emission in turbulence mostly concentrates near magnetic islands. The emission is stronger toward the edges than at the centers of islands, consistent with the nonthermal particle distributions. As a consequence, the local polarized intensity also appears higher toward the boundaries of magnetic islands. However, the directions of the polarization vectors are apparently random: they can be parallel, perpendicular, or oblique to the edges of islands. This explains the overall low PD in Figure \ref{fig:Case1temp}, as local polarized intensity with different PA orientations can vector-average to cancel most of the total polarization. Very few local regions dominate the polarized flux over other regions, and there is no overall trend in the polarization vectors across individual cells. Consequently, the PD and PA only undergo small fluctuations (see Figure \ref{fig:Case1temp}), which can be approximated by random walks about some mean values.

One may note that the emission map appears to cover a larger region than the nonthermal particle map. This is not caused by the relatively low resolution of the radiative transfer simulation compared to the PIC grid. Rather, it occurs simply because somewhat lower-energy particles contribute modestly to the optical emission, which occupies regions closer to the centers of islands. This is more apparent in Figure \ref{fig:Case1mapmultiband}, where we plot the corresponding nonthermal particle density and polarized emission maps at infrared and ultraviolet bands. Comparison of the first and third rows in Figure \ref{fig:Case1mapmultiband} reveals that the higher-energy electrons have overall lower density, and are distributed mostly in small regions at the edges of magnetic islands. Therefore, their emission represents the regions with the strongest particle acceleration, which results in more pronounced variability. Additionally, since higher-energy electrons only exist in small regions, there is less chance that the local polarized flux will be canceled out by other regions with different PA orientations. Therefore, the higher-energy bands are expected to have higher mean values of PD from synchrotron radiation. Nonetheless, this effect is unclear between the infrared and ultraviolet bands in our simulations, because the two bands are only separated by a small factor in $\gamma_e$. We expect a larger and detectable difference between optical and X-ray polarization of high-frequency-peaked blazars, which will be presented in a future work.

\section{Polarization Angle Swings in Turbulent Plasma\label{sec:PAswings}}

The time-dependent radiation signatures from relativistic turbulence depend on the initial wave phases, which we treat as a set of random numbers in cases 1-3. This is because different initial phases can lead to different spatial distributions of turbulent structures and nonthermal particles. In certain situations, PA swings can appear during the evolution of turbulence. The swings can be either smooth or irregular, during which the PD drops to nearly zero. These swings are apparently due to random walks instead of deterministic processes. The PA swings do not necessarily coincide with the peak of a flare. Here we show two cases, 2 and 3, that exhibit smooth PA swings. Interestingly, Case 3 represents a very rare situation where an orphan optical PA swing happens without a counterpart at either infrared or ultraviolet bands.

\subsection{Multi-Wavelength PA Swings}

\begin{figure}
\centering
\includegraphics[width=\linewidth]{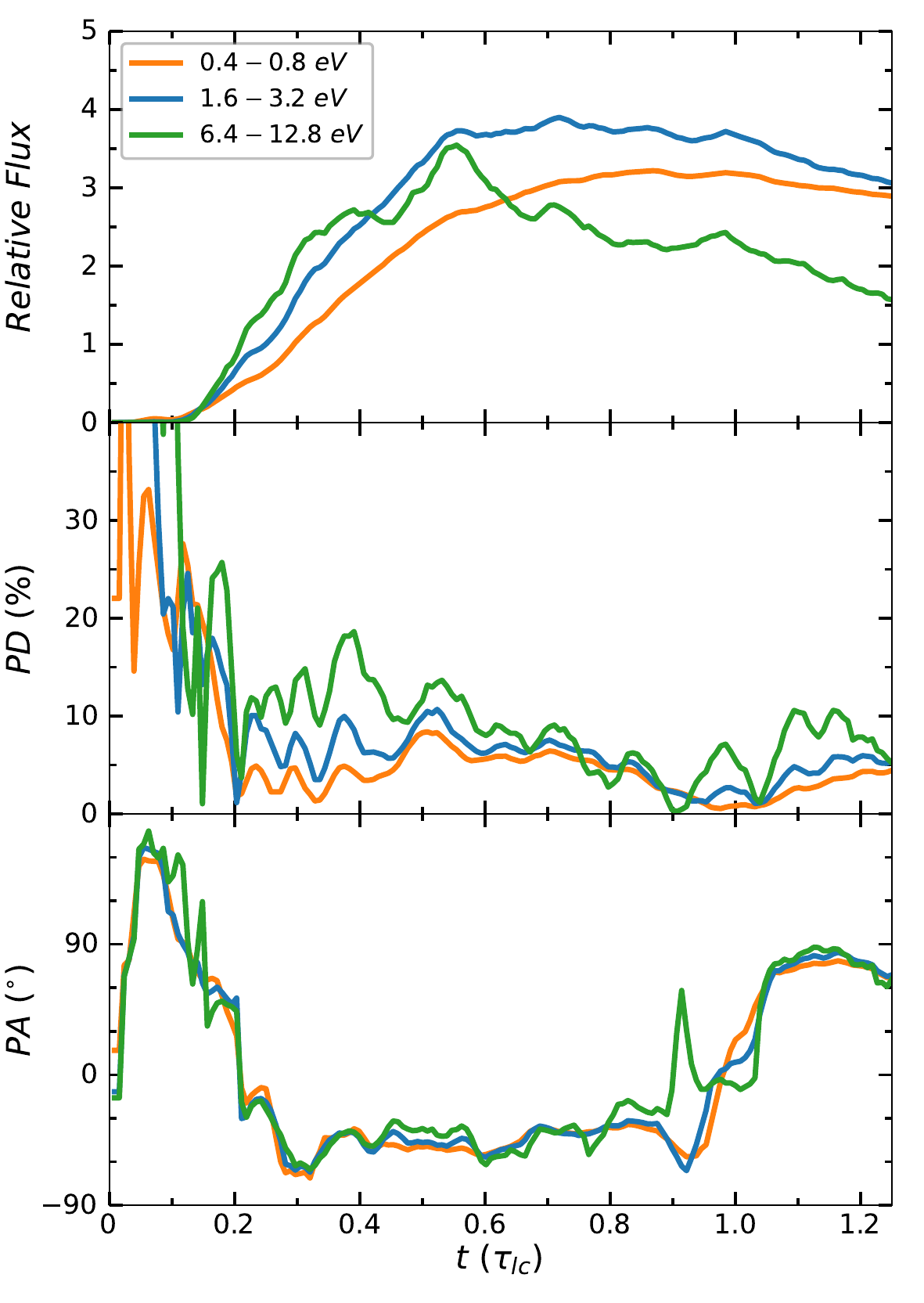}
\caption{Same as Figure \ref{fig:Case1temp} but for Case 2.}
\label{fig:Case2temp}
\end{figure}

\begin{figure*}
\centering
\includegraphics[width=\linewidth]{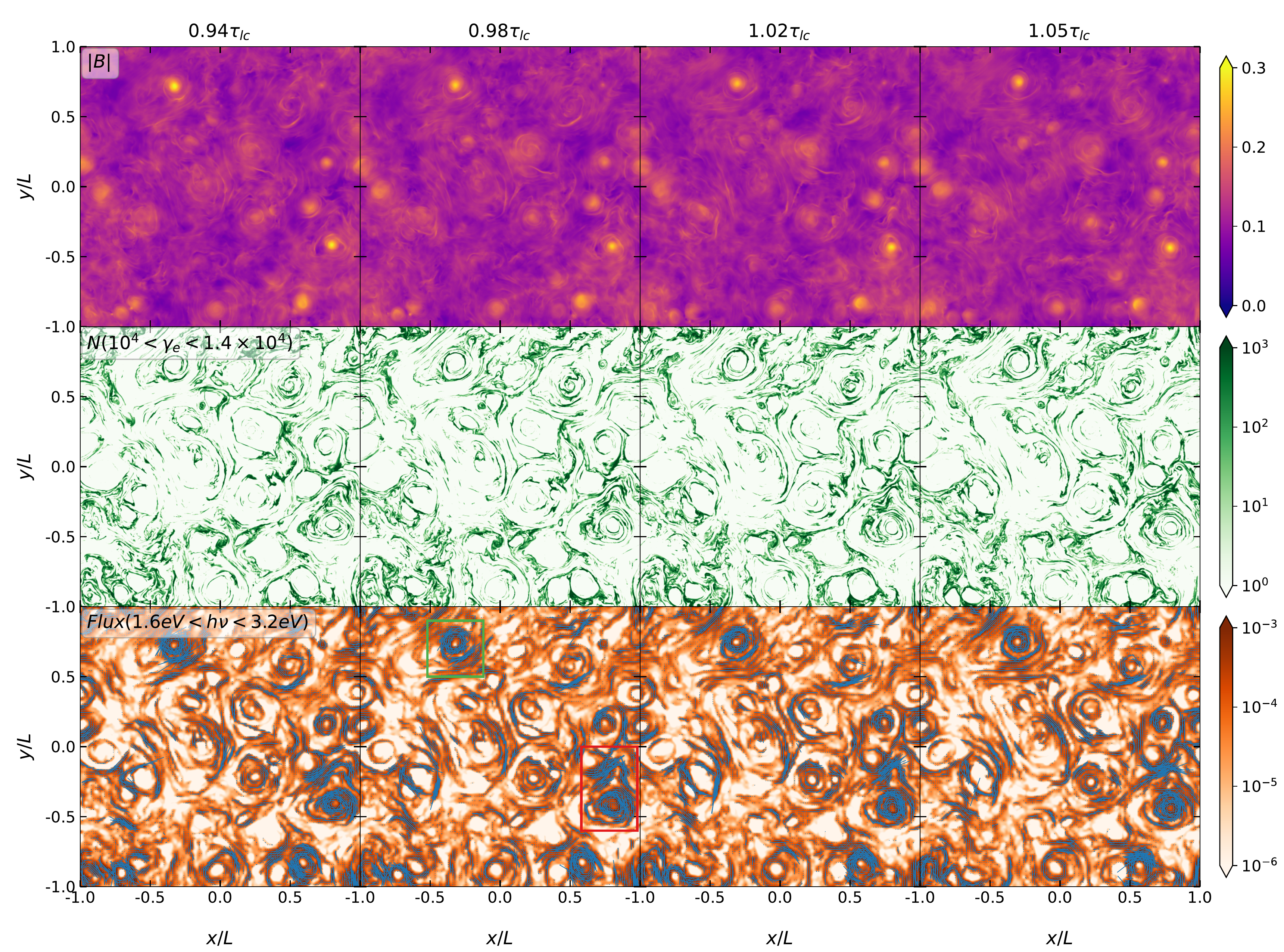}
\caption{Snapshots of the magnetic field, nonthermal particle density, and optical polarized intensity map during the PA swing in Case 2. Snapshots are plotted similarly to Figure \ref{fig:Case1map}. The green and red boxes in the last row, second column highlights a magnetic island and an interaction region between two magnetic islands, respectively.}
\label{fig:Case2mapPAswing}
\end{figure*}

While Case 1 represents the majority of initial phase realizations that only show small fluctuations of polarization signatures, Case 2 represents a few runs that show PA swings. Its only difference from Case 1 is that it starts with a different set of initial random phases. Therefore, we can see that the multi-wavelength light curves and PDs are similar to Case 1 (Figure \ref{fig:Case2temp}). The spectral evolution for both particles and photons is almost identical to Case 1, so we do not show it here. The key difference here is that Case 2 has a $\sim 150^{\circ}$ swing near the end of the simulation (from $0.9\tau_{lc}$ to $1.1\tau_{lc}$). The swing looks rather smooth, without any spikes, and is apparently associated with the start of a decrease in optical flux. We note that, unlike the other two bands, in the ultraviolet band a spike occurs in the PA evolution at $t\sim 0.9\tau_{lc}$. But, as one can see in the PD curve, the ultraviolet band has zero PD at that time. Therefore, this spike in PA results from the PA being arbitrary when the PD is zero. The rather smooth PA swings appear to imply a deterministic origin. However, after examining four snapshots of the simulation during the PA swing (Figure \ref{fig:Case2mapPAswing}), we do not observe any clear difference from Figure \ref{fig:Case1map}: the emission is still mainly around the magnetic islands, and the local polarized intensity is stronger at the edge of these islands. Similar to the last row in Figure \ref{fig:Case1map}, there are several regions with more polarized flux; however, by isolating regions with either magnetic islands (such as the green box) or interactions between islands (such as the red box), we find that none of them dominate the total polarization evolution or show a similar PA swing. We do not observe any systematic patterns in the magnetic field or nonthermal electron distributions in Figure \ref{fig:Case2mapPAswing} upper and middle rows either. Additionally, since the multi-wavelength light curves are already decreasing without major fluctuations during this epoch, we conclude that the PA swings are of a stochastic nature. Therefore, one expects that PA swings from turbulence can have arbitrary amplitudes, but due to the stochastic nature, large-amplitude swings should be rare.

\subsection{Orphan PA Swing}

\begin{figure}
\centering
\includegraphics[width=\linewidth]{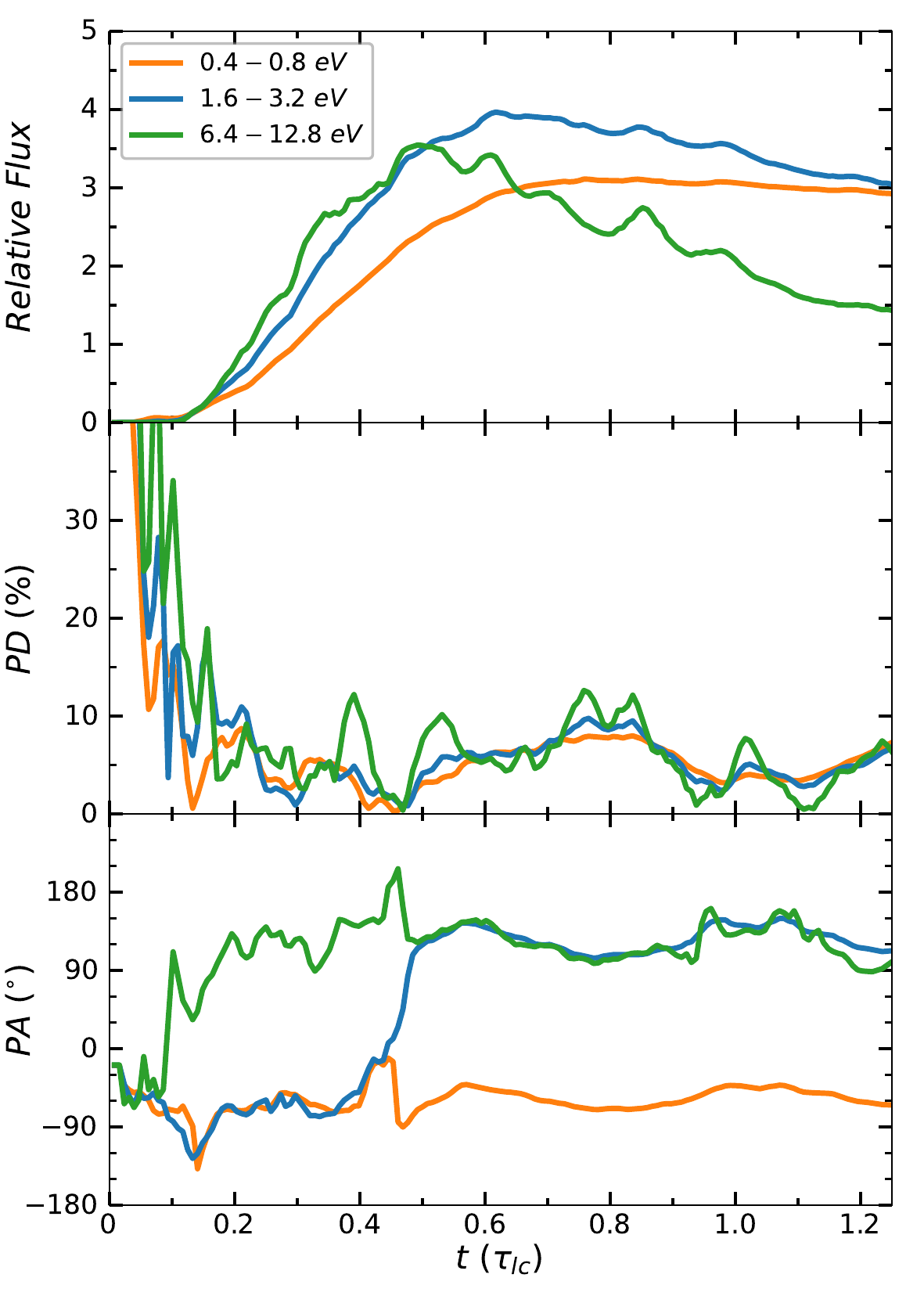}
\caption{Same as Figure \ref{fig:Case1temp} but for Case 3.}
\label{fig:Case3temp}
\end{figure}

Case 3 is the only one that includes an orphan optical PA swing without infrared or ultraviolet counterparts. As shown in Figure \ref{fig:Case3temp}, the overall light curves and PDs are similar to Cases 1 and 2. The PA curves of the three bands are generally similar, except that the optical band exhibits a swing of $\sim 210^{\circ}$, while in the other two bands there is a $\lesssim 60^{\circ}$ change in the opposite direction. This optical swing is also the largest smooth PA rotation over all our simulations. However, the PD curves appear very similar for the three bands, except that the ultraviolet PD is slightly more variable owing to the stronger cooling. We note that the PDs of infrared and ultraviolet bands drop to zero momentarily during this epoch, but the optical PD remains at a few percent during the swing. Thus the optical PA swing is not due to an arbitrary value of PA when PD reaches zero. The swing is not linked to the optical flare peak, either: it takes place during the rising phase of the flare. Since the magnetic field evolution is the same for the three bands, this suggests that the spatial distributions of nonthermal particles of different Lorentz factors are also stochastic and lead to the orphan PA swing.

\section{Correlation Length and Polarization} \label{sec:correlationlength}

Our studies of the three representative runs clearly show that the radiation signatures from turbulence are of a stochastic nature. However, turbulent structures can appear on both large and small physical scales. It is important to understand at which scale structures dominate the radiation and polarization patterns. Here we perform additional analyses: we examine the effects of the resolution of the radiative transfer simulations, PIC simulation box size, and injection of magnetic fluctuations. Our results suggest that the variability and average PD are governed by the turbulence correlation length. Specifically, light curves and polarization signatures are more strongly variable with smaller correlation length, and the average PD is proportional to the ratio of correlation length to box size. All parameter studies in this and the next section adopt the same initial phase realization as Case 2 in the previous section. We choose Case 2 because it shows PA swings in three bands, which are very sensitive to the magnetic field morphology and evolution. This can help us to detect how different physical parameters can affect the spatial distributions of magnetic field and nonthermal particles.

\subsection{Effects of Radiative Transfer Resolution}

\begin{figure}
\centering
\includegraphics[width=\linewidth]{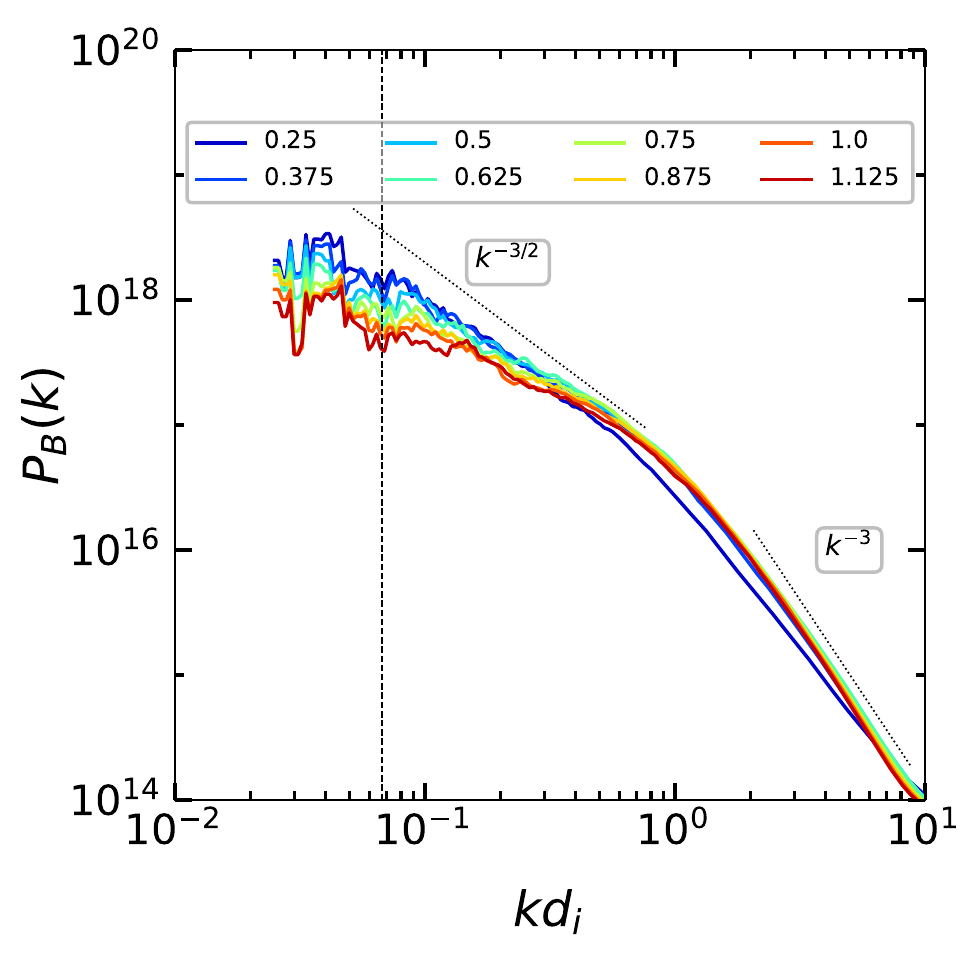}
\caption{Snapshots of the power spectra of the magnetic field for Case 2. The black vertical line marks the correlation length.}
\label{fig:Case2BPS}
\end{figure}

\begin{figure}
\centering
\includegraphics[width=\linewidth]{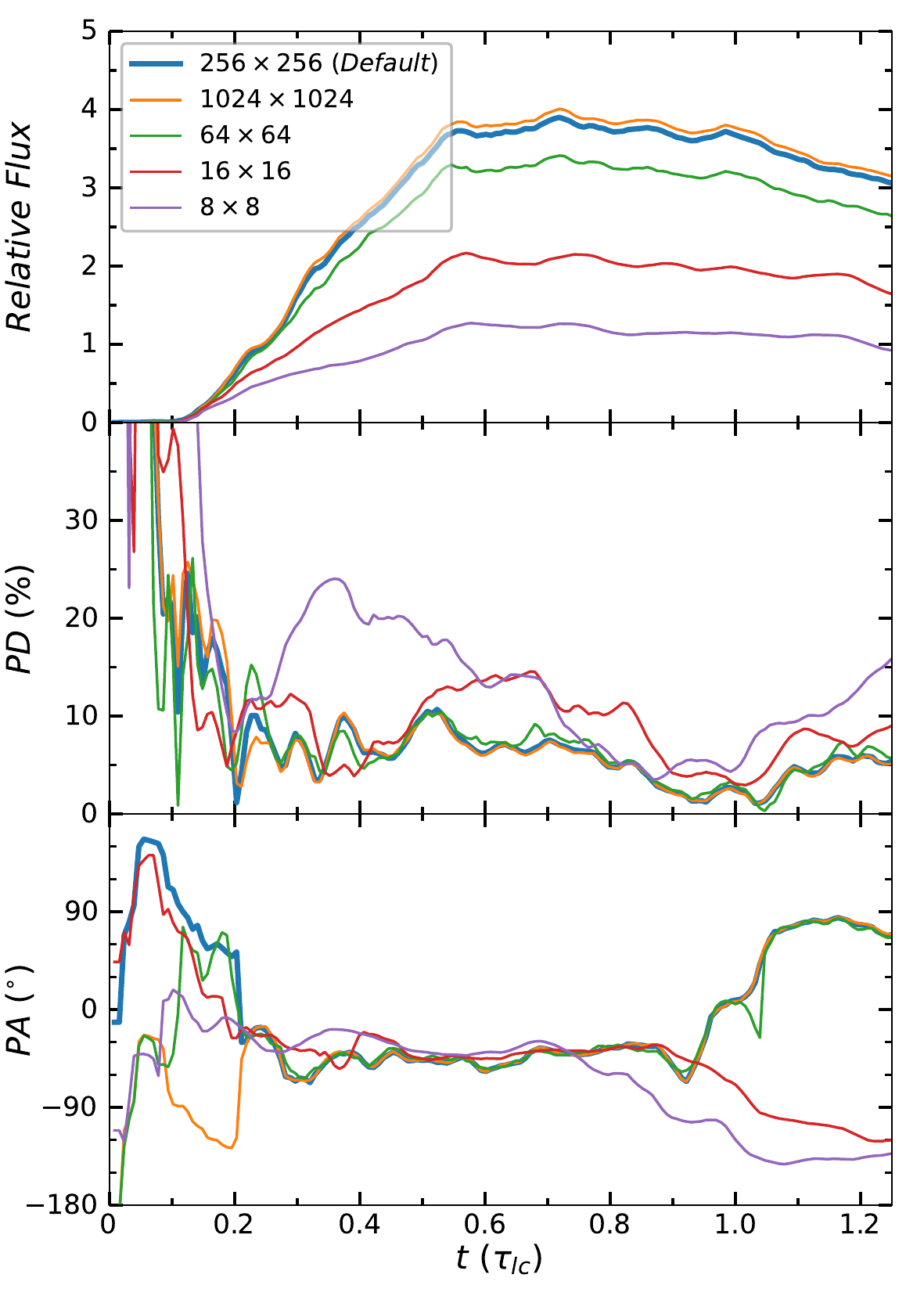}
\caption{Optical light curves, PDs, and PAs for different radiative transfer grids for Case 2.}
\label{fig:Case2resolution}
\end{figure}

Figure \ref{fig:Case2BPS} plots the power spectra of the magnetic field for Case 2. It is apparent that the spectra have the form of a broken power-law, with a break at the ion kinetic scale, $\sim d_i$. This is consistent with \citet{Comisso2018}, and may be due to strong kinetic dissipation. The black vertical line marks the start of the broken power-law, which represents the correlation length of the turbulence, approximately equal to the wavelength of the largest wave vector among the initial random phases. To test at what physical scales the turbulent structures have the strongest impact on radiation, we perform radiative transfer simulations with different resolutions, which range from the kinetic scale $d_i$ ($1024\times 1024$ grids) to one leaving the correlation length unresolved ($8\times 8$ grids). Figure \ref{fig:Case2resolution} plots the optical light curves and polarization evolution with different resolutions. We remind the reader that in the default grid size, we sum $32\times 32$ PIC cells to obtain one radiative transfer cell, so that we have very good statistics for particle spectra in each cell. But upon averaging the magnetic field in $32\times 32$ PIC cells, the magnetic field in the radiative transfer cell is completely ordered, hence any disorder of the magnetic field on a scale smaller than the radiation transfer grid size is ignored. We find that, as long as the correlation length is well resolved (equal to or more than $64\times64$ grids in Figure \ref{fig:Case2resolution}), the temporal behaviors of the light curve and polarization are nearly identical after the turbulence fully develops at $t\sim 0.2\tau_{lc}$. The total flux decreases with coarser resolution because, by averaging the magnetic field, some turbulent magnetic field components in opposite directions may cancel out, resulting in a lower magnetic field compared to the original value. Therefore, we suggest that the radiation signatures from turbulence are dominated by turbulent structures at the correlation length. The kinetic-scale dynamics, although important for plasma evolution and particle acceleration, have minimal effects on the observed properties of the radiation.

\subsection{Effects of Correlation Length}

\begin{figure}
\centering
\includegraphics[width=\linewidth]{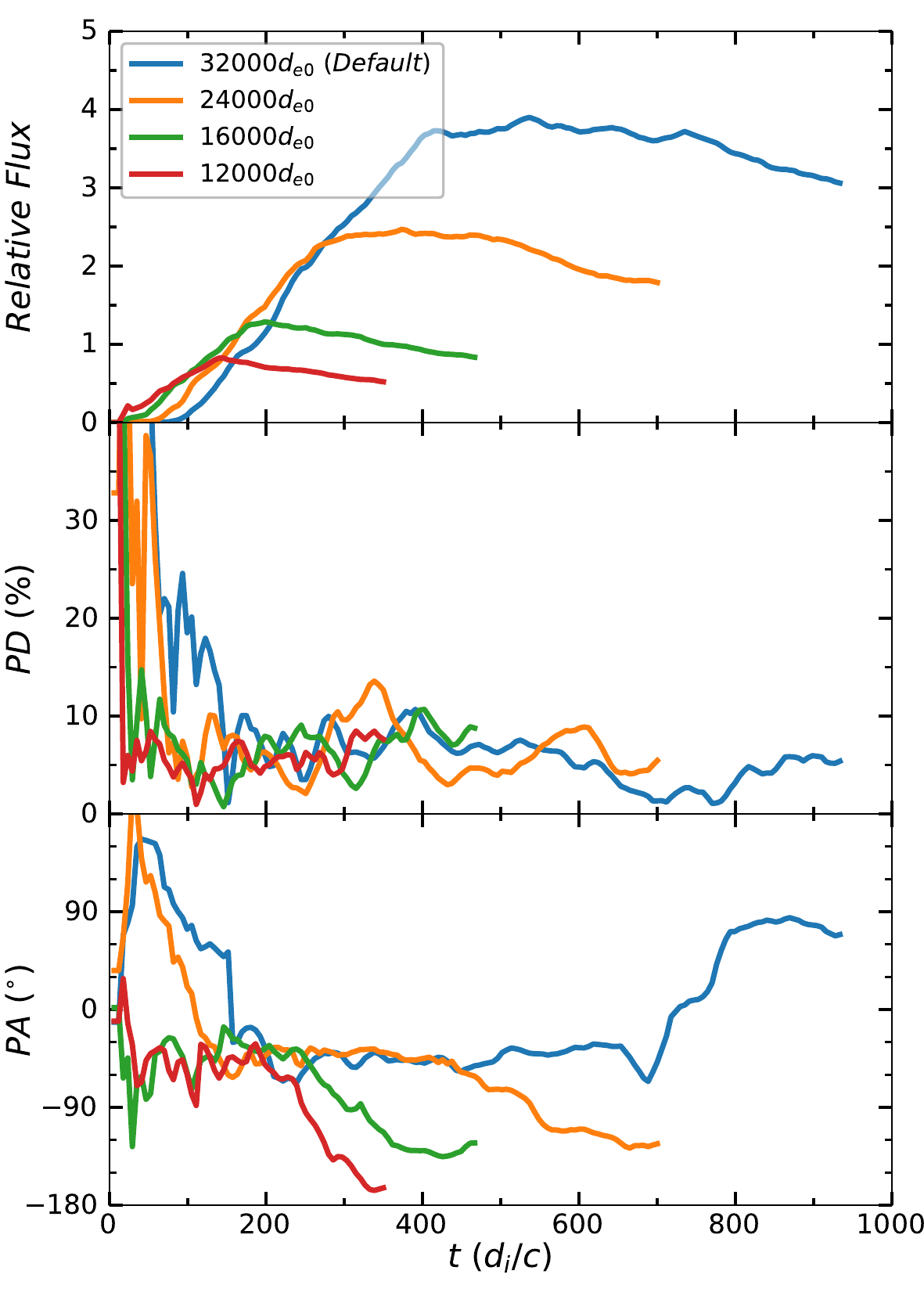}
\caption{Optical light curves, PDs, and PAs for different simulation box sizes. Time is in units of $d_i/c$.}
\label{fig:Case2boxsize1}
\end{figure}

\begin{figure}
\centering
\includegraphics[width=\linewidth]{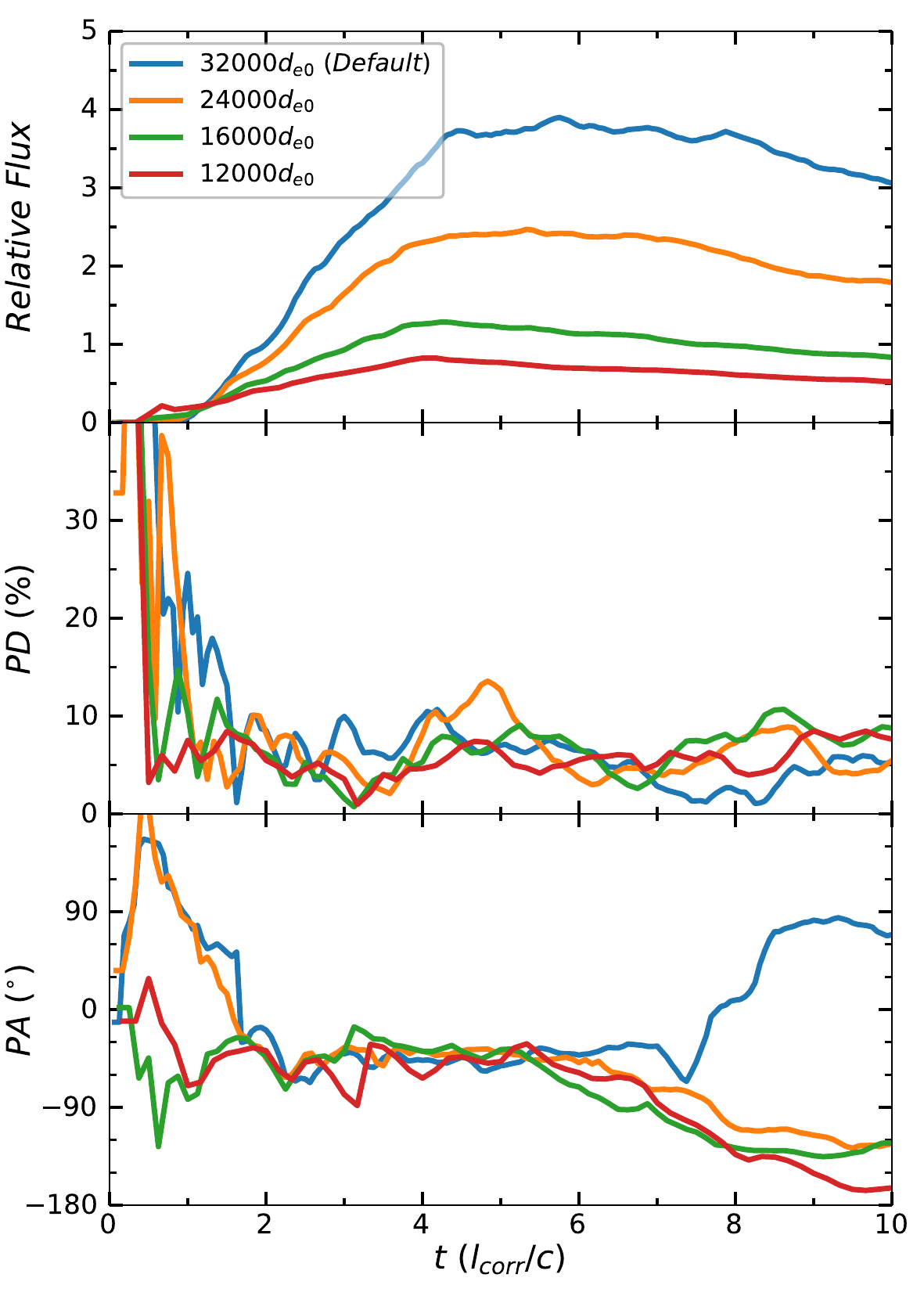}
\caption{Optical light curves, PDs, and PAs for different simulation box sizes. Time is in units of $l_{corr}/c$.}
\label{fig:Case2boxsize2}
\end{figure}

Figure \ref{fig:Case2boxsize1} compares the optical light curves and polarization signatures for different simulation box sizes. It is clear that the rising slope of the light curve is nearly identical for different box sizes, indicating that the overall particle acceleration rate is similar. As can be seen, both PD and PA vary faster, with shorter flare duration, as the box size is decreased. However, the average PD variations appear to depend weakly on box size. All runs include PA swings at the end of the simulation, though the direction and amplitude of rotation can vary. We have tried to identify any deterministic patterns during the PA swings for smaller box simulations similar to Figure \ref{fig:Case2mapPAswing}, but none is found, as expected, suggesting a stochastic origin similar to Case 2. If we plot the temporal evolution in units of light crossing time of the correlation length of each run, we find that, except for the maximum flux and PA rotation direction, the variations are of a stochastic origin. The general evolution of light curves and polarization signatures are nearly identical for the four runs (Figure \ref{fig:Case2boxsize2}).

\begin{figure}
\centering
\includegraphics[width=\linewidth]{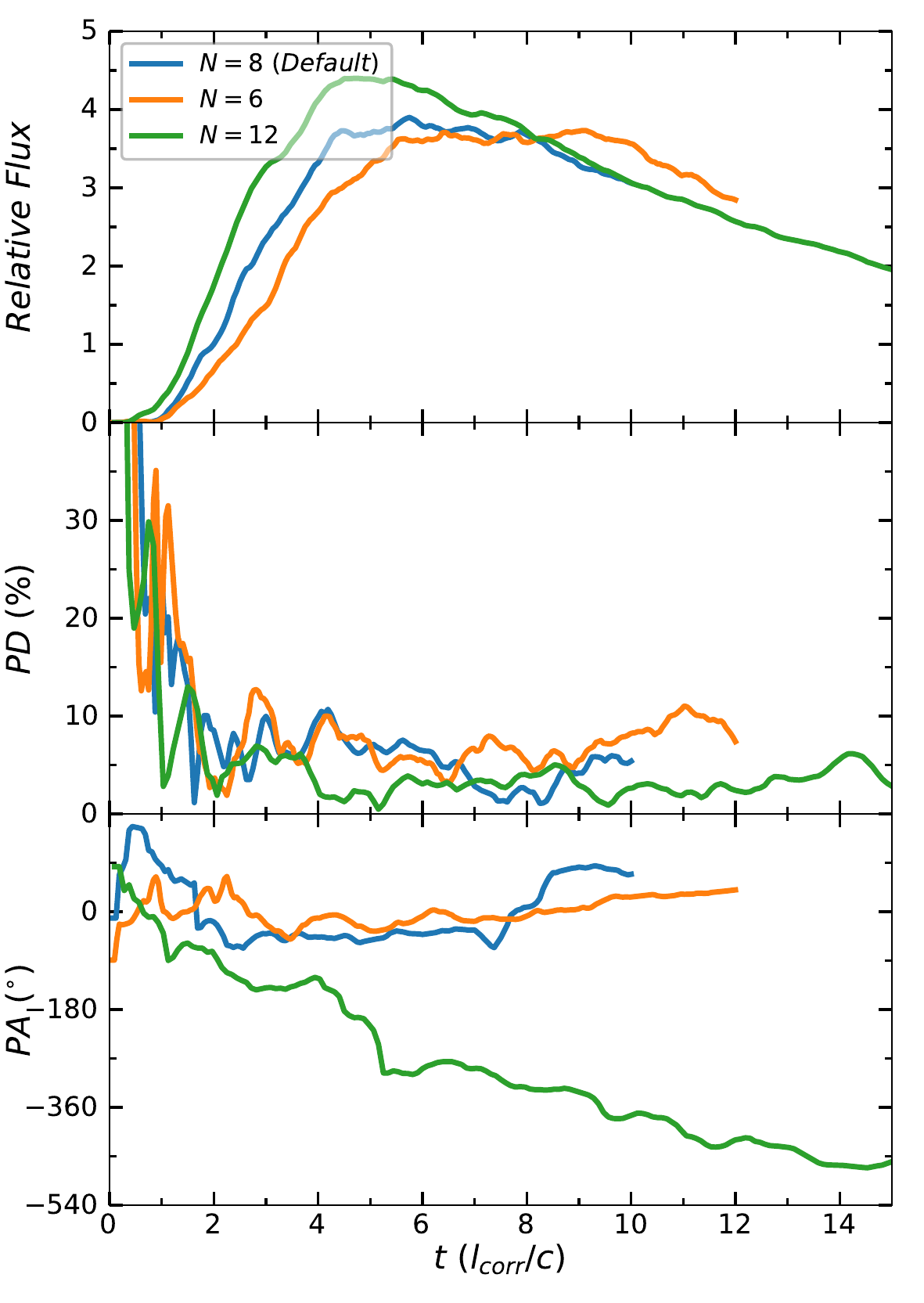}
\caption{Optical light curves, PDs, and PAs for different mode numbers of the initial magnetic fluctuations. Time is in units of $l_{corr}/c$.}
\label{fig:Case2nmodes2}
\end{figure}

Inspired by the above findings, we examine whether the ratio of the correlation length to the box size can affect the radiation signatures. This ratio can vary by changing the number of modes in the initial magnetic fluctuations. Figure \ref{fig:Case2nmodes2} shows the comparison. Here the simulation box size is the same for these three runs, thus the larger mode number means that the correlation length is smaller. Therefore, we observe that the smaller correlation length leads to an earlier peak in the light curve, indicating that the energy dissipation happens earlier. Most importantly, we find that after the turbulence has developed, the average PDs of the three runs are proportional to the correlation length. This is expected, because the magnetic field morphology can be considered as random for physical scales larger than the correlation length. Under this scenario, the PD is proportional to $1/\sqrt{N}$ \citep{Burn1966}, where $N$ is the number of cells that have independent random magnetic field. Since our simulation is 2D, $N=S/S_0$, where $S=4L^2$ is the size of the simulation box, and $S_0=l_{corr}^2$, then we have $PD\propto l_{corr}/(2L)$.

\section{Additional Parameter Studies \label{sec:parameterstudy}}

\begin{figure}
\centering
\includegraphics[width=\linewidth]{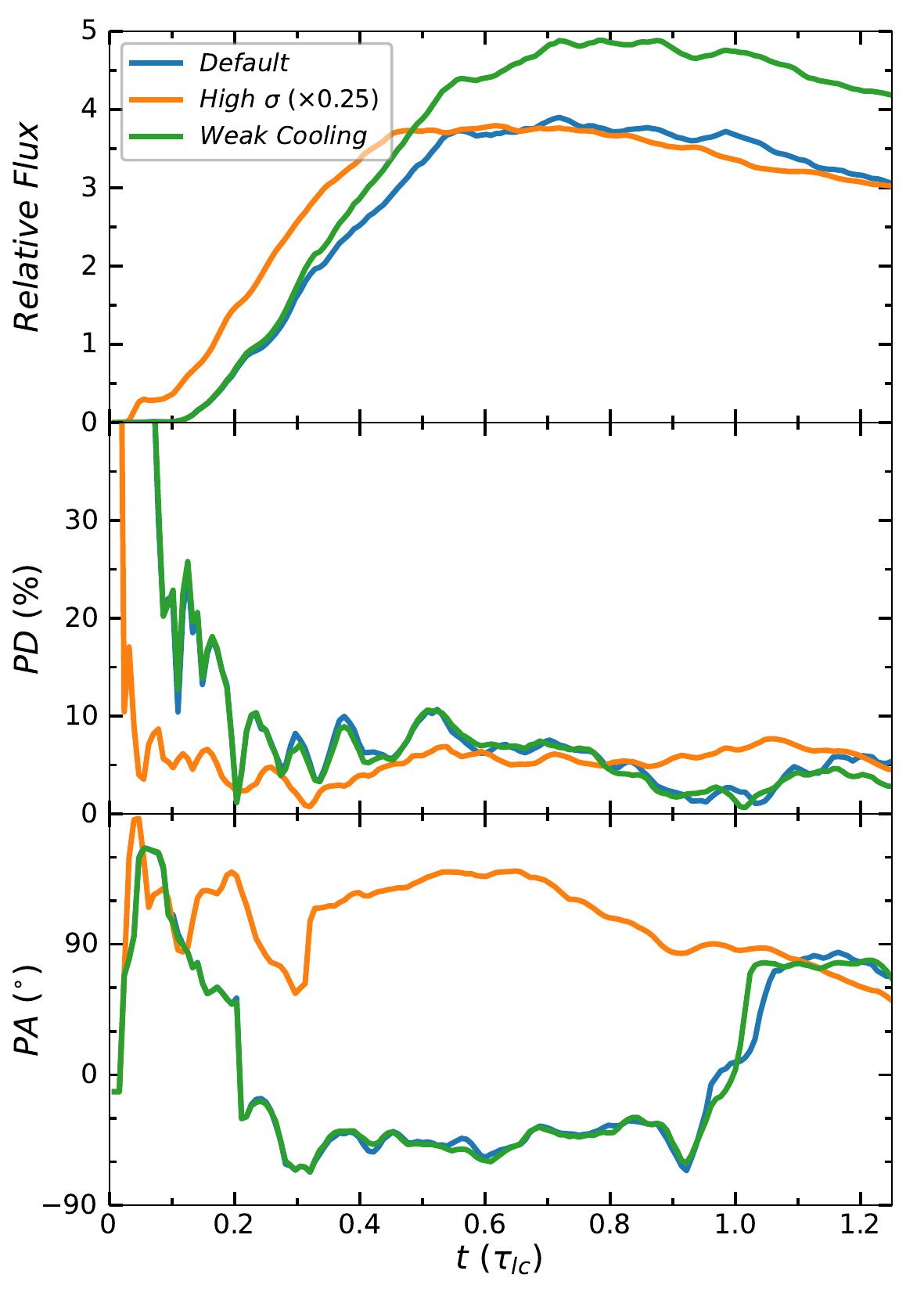}
\caption{Similar to Figure \ref{fig:Case2resolution}, but for optical emision from the default Case 2 run, another run with higher $\sigma$, and a third run with weaker cooling. The relative flux for the high $\sigma$ run is multiplied by a factor of 0.25 to be shown in the same plot range.}
\label{fig:Case2sigmacooling}
\end{figure}

Here we explore the effects of magnetization and cooling factors on radiation and polarization signatures. Both parameters are important to the particle spectral evolution, but apparently they only weakly affect the general temporal behavior. Figure \ref{fig:Case2sigmacooling} presents a comparison between Case 2 and a high-$\sigma$ run, as well as a run with weak cooling. The high-$\sigma$ run has a magnetization factor $\sigma$ four times that of the default run. We still normalize the average initial magnetic field strength to $0.1~\rm{G}$. The weak cooling run has a radiative cooling reaction force two times lower than does the Case 2 setup. The high-$\sigma$ case peaks earlier due to stronger particle acceleration. The weak cooling case, on the other hand, peaks later than the default Case 2 light curve, followed by a longer cooling tail. Weak radiative cooling apparently has trivial effects on polarization signatures. This is expected, since the weaker cooling essentially pulls the optical band below the spectral cooling break, making it similar to the infrared band, which has nearly identical light curves and polarization evolution as in Case 2. It is worth noting that the high-$\sigma$ case has a long-term PA swing from $t\sim 0.65\tau_{lc}$ to the end of the simulation, with an amplitude of $\sim 120^{\circ}$. It does not correlate with the peak of the light curve or a drop in the PD. We have performed the same analyses as in the previous section to explore deterministic processes, but none is found. Therefore, this long-term PA swing should also be attributed to the stochastic evolution of the large-scale turbulence.

\section{Comparison with TEMZ Simulations \label{sec:temz}}

The TEMZ model \citep{Marscher2014}, which is designed to simulate emission from turbulent plasma crossing a standing conical shock in a relativistic jet, approximates the effects of turbulence via the scheme of \citet{Jones1988}. The numerical code divides the volume beyond the shock into thousands of turbulent cells. Every cell belongs to four nested turbulent ``zones,'' each with its own (randomly selected) uniform magnetic field and density of relativistic electrons. The smallest zone consists of the cell itself and the others contain $2\times2\times2$, $4\times4\times4$, and $8\times8\times8$ cells, the last of which encompasses 1/4 of the jet cross-section. The magnetic field of a cell is the vector sum of the field in all of the cells' zones, weighted by zone size according to the Kolmogorov spectrum. The density of a cell is calculated in a similar manner. The turbulent rotational velocity of a cell is the weighted vector sum of the four-velocities of the cells' zones, with the center of the respective zone serving as the center of rotation in order to simulate vortices. The turbulent velocity is added relativistically to the laminar velocity of the jet flow beyond the shock.

After gaining energy as they cross the shock front, the electrons lose energy from synchrotron and Compton radiation. The volume filling factor is therefore lower for higher-energy electrons, which are present only in a thin layer beyond the shock front. This smaller volume leads to more pronounced variability, as well as polarization that has a higher mean and is more rapidly variable
at higher frequencies. It also limits the higher-frequency flux so that the spectral break is $>0.5$, as is generally observed in spectral energy distributions \citep{Abdo2010}.

\begin{figure}
\centering
\vspace{-0.2cm}
\includegraphics[width=0.8\linewidth]{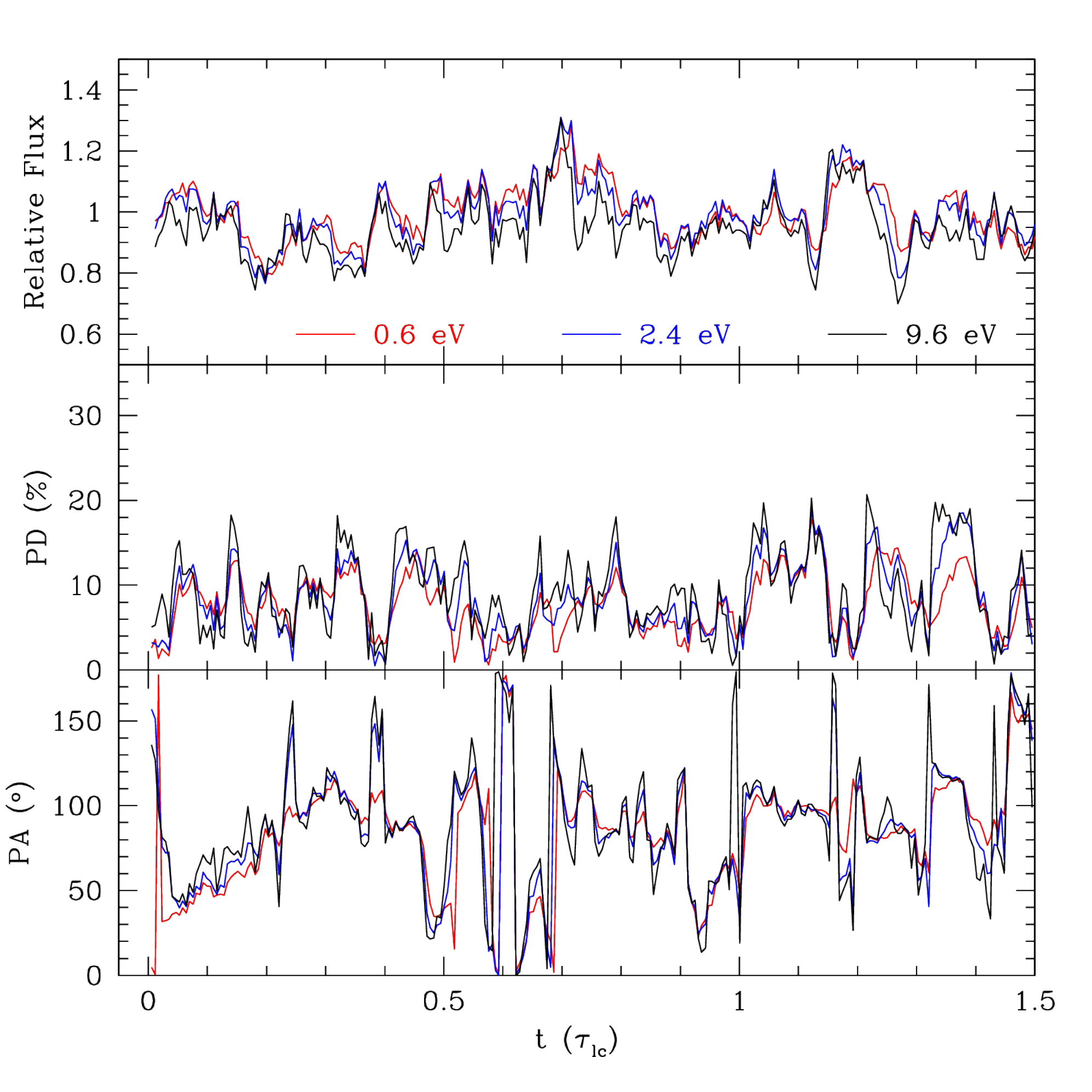}
\vspace{-1cm}
\includegraphics[width=0.8\linewidth]{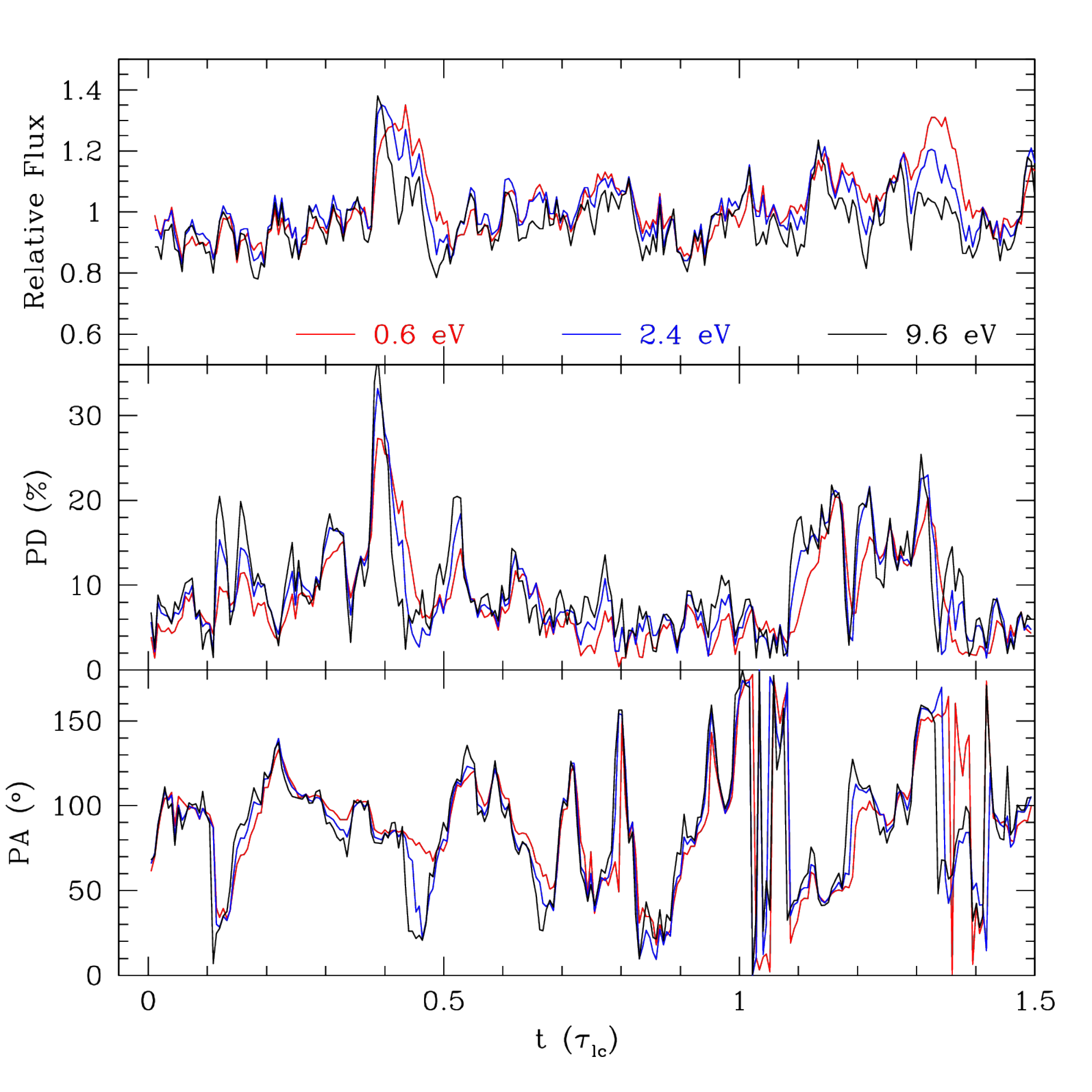}
\vspace{-1cm}
\includegraphics[width=0.8\linewidth]{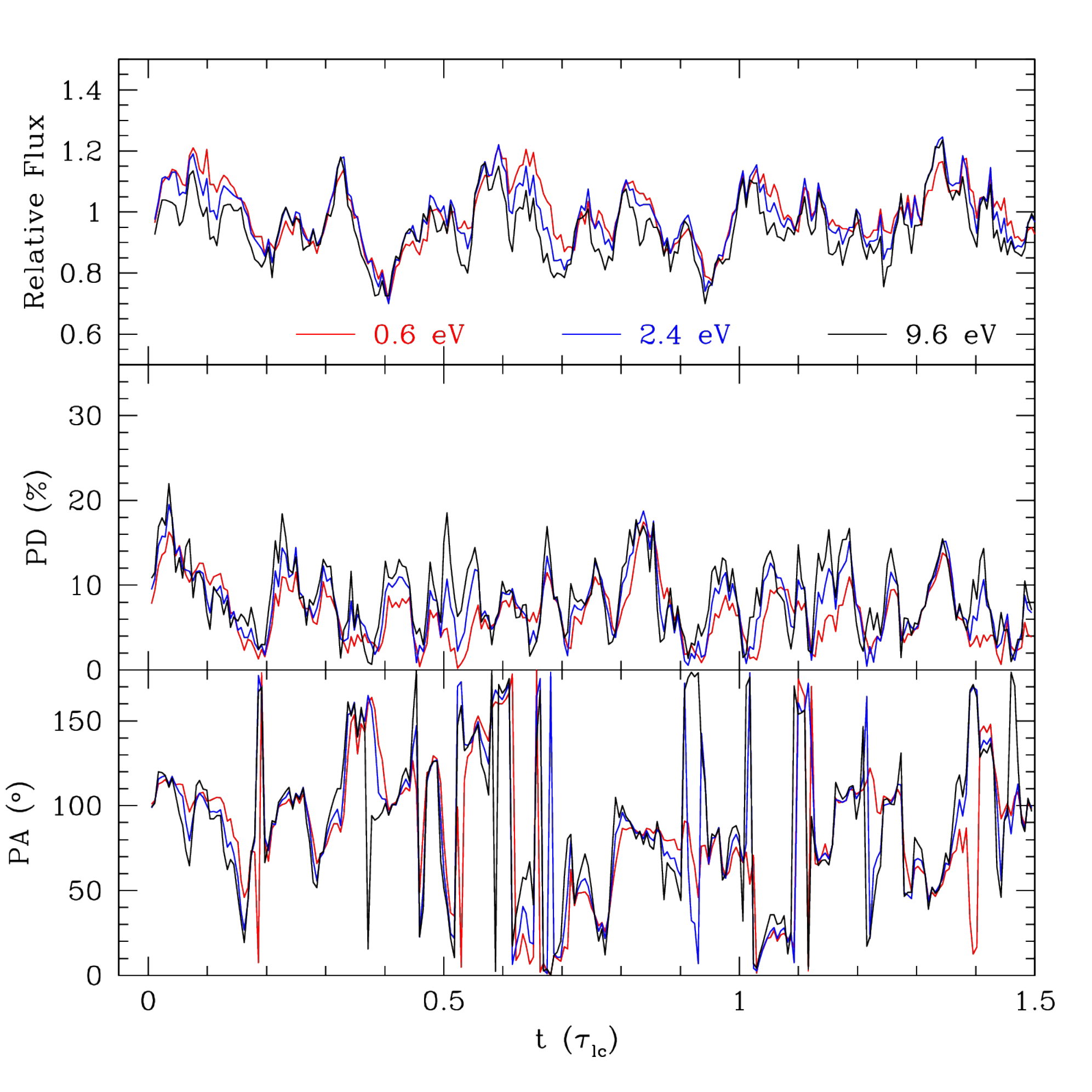}
\caption{Sample simulation of flux and polarization vs.\ time of
the TEMZ model over three time ranges, each spanning 1.5 light-crossing times, for comparison with Figures \ref{fig:Case1temp}, \ref{fig:Case2temp}, and \ref{fig:Case3temp}. The total simulation extends over 125 light-crossing times.}
\label{fig:temzlc}
\end{figure}

\begin{figure}
\centering
\vspace{-0.2cm}
\includegraphics[width=\linewidth]{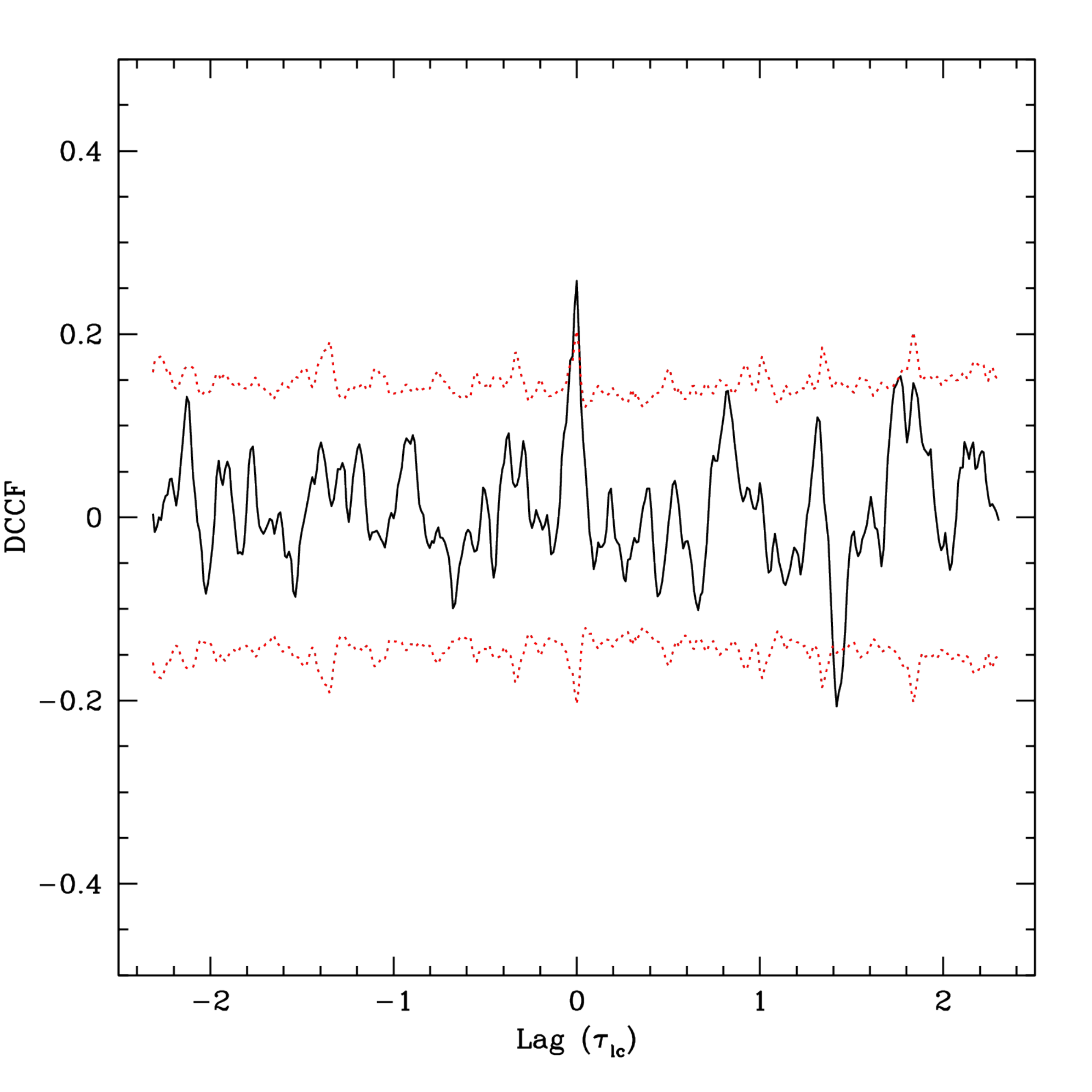}
\vspace{-2.6cm}
\caption{Solid line: discrete cross-correlation function \citep[DCCF;][]{Edelson1988} optical (2.4 eV) flux and polarization of the TEMZ simulation presented in Figure \ref{fig:temzlc} over the full 125 light-crossing times. Positive lag corresponds to variations in flux leading variations in polarization. Red dotted lines: $\pm3\sigma$ significance level.}
\label{fig:temzdccf}
\end{figure}

The simulated TEMZ flux and polarization curves, examples of which (three time intervals from the same simulation) are displayed in Figure \ref{fig:temzlc}, provide an example of the variability under the model. There is no initial increase in flux because the simulated turbulence is fully set up from the beginning. The general patterns are similar with our combined PIC and polarized radiative transfer simulations, including the relatively small-amplitude flux variations and low PD. The PA has no preferred orientation, and can occasionally undergo swings. None of the swings in Figure \ref{fig:temzlc} is larger than $180^{\circ}$. As shown in Figure \ref{fig:temzdccf}, flux and polarization appear uncorrelated, except for a weak, but significant, correlation at zero lag, as well as an apparently spurious anti-correlation at a similar significance level at a lag of $+1.4\tau_{\rm lc}$. Visual inspection of the middle portion of Figure \ref{fig:temzlc} suggests that the correlation is due to occasional major flares. The flux and polarization could increase together if the flare arises from a region where a particularly high magnetic field strength or relativistic electron density occurs over a small volume with nearly uniform magnetic field direction. 

The main difference between the TEMZ and PIC simulations is that fluctuations of flux, PD, and PA in TEMZ are stronger than in the combined PIC and polarized radiative transfer simulation. There are several reasons for this. First, the turbulent cells in TEMZ cross the shock at a highly relativistic speed, which effectively changes rapidly the density and magnetic field of the regions that emit at high frequencies. In contrast, the PIC turbulence changes more slowly, since there are no relative motions near the speed of light. Also, the PIC simulation uses periodic boundary conditions, so that no plasma or energy is injected into the simulation domain. The TEMZ model, on the other hand, allows plasma to flow in and out of the simulation domain. Second, the standing shock in the illustrated TEMZ simulation has a cone shape, which decreases the coherence of the turbulent zones. A moving shock with shock normal along the jet direction would have more coherence of adjacent emitting cells that belong to the same nested turbulent zones. Most of the radiating particles in the PIC simulations are in fairly long filaments. Finally, the PIC simulation is 2D, which only allows us to track the magnetic field variations in the simulation plane. TEMZ, on the other hand, is 3D, including evolution in the third direction, which can lead to additional variations.

\section{Implications for Observations \label{sec:implication}}

Turbulence is ubiquitous in astrophysical systems. In blazar jets, turbulence may be driven by shock and/or magnetic instabilities, which can accelerate nonthermal particles by dissipating kinetic or magnetic energy, respectively. Recent multi-wavelength blazar observations, including optical polarization monitoring programs, have revealed apparent stochastic patterns in blazar emission, which are strong evidence for turbulence \citep{Marscher2021}. Our combined PIC and polarized radiative transfer simulations of turbulence in the blazar zone environment systematically study, for the first time, synchrotron radiation and polarization signatures from first principles. Our results confirm that turbulence results in overall stochastic radiation patterns. Spectra and light curves resulting from turbulence are roughly consistent with other mechanisms, such as shocks and magnetic reconnection. The best chance to distinguish turbulence from other mechanisms is to analyze multi-wavelength polarization.

Turbulence leads to a harder-when-brighter trend due to cooling. The result is a broken power-law blazar spectrum, with spectral indices ranging from $2\lesssim p\lesssim 4$, as shown in our simulations. These results are consistent with typical blazar observations, as well as with the predictions from other models, although magnetic reconnection can produce spectral indices harder than 2 \citep{Guo2014,Sironi2014}. Multi-wavelength light curves are well correlated, and higher-frequency bands appear more variable. No high-amplitude (greater than a factor of $\sim2$), rapid variability in light curves is found in our simulations, but this is because our choice of parameters does not generate considerable anisotropic particle distributions. If the nonthermal particles are anisotropic in certain local structures \citep{Comisso2021}, or the local Lorentz factors vary significantly (as shown in the TEMZ simulation), turbulence can give rise to fast variability. Variability in light curves is mostly due to nonthermal particle evolution, thus in a leptonic model we expect correlated high-energy variability from Compton scattering. Our simulations also find that turbulence can equally efficiently accelerate electrons and protons, thus even in a hadronic model, the high-energy variability is expected to be associated with the synchrotron component, except for the difference in electron and proton cooling \citep{Zhang2016b}.

Both PD and PA tend to fluctuate around a mean value in a turbulent blazar zone. The average PD depends on the correlation length of the turbulence. There is no preferred PA orientation for turbulence. Recent radio observations have found that many blazars (especially flat spectrum radio quasars) do not have preferred PA orientation compared to the jet direction, which may be a signature of turbulence \citep{Hodge2018,Marscher2021}. Turbulence can produce significant PA rotations, although, due to its stochastic nature, such events are not common. The amplitude, duration, and direction of the PA swings are arbitrary in a turbulent environment, but large-amplitude ($\gtrsim 180^{\circ}$) swings occur infrequently in our simulations. \citet{Kiehlmann2017} have suggested that deterministic PA swings should appear smoother, which can help to distinguish turbulent PA swings from shock and/or magnetic reconnection. However, at least some PA swings from our simulations appear very smooth. Therefore, we suggest that ``smoothness'' is not a robust diagnostic for turbulence. Instead, PA swings are not physically connected to multi-wavelength flares or drop of PD in turbulence. Previous works have found that many optical PA swings are associated with flares, and the PD tends to decrease during such events \citep{Blinov2018,Marscher2010,Chandra2015}, thus disfavoring a turbulent origin.

\section{Summary and Discussion \label{sec:discussion}}

This paper presents a systematic study of the synchrotron radiation and polarization signatures from turbulence in the blazar zone environment. Our approach combines PIC and polarized radiative transfer simulations to self-consistently study radiation signatures. We have clearly demonstrated that the radiation patterns are of stochastic origin, as expected from turbulence, and the key parameter that governs the radiation is the correlation length of the turbulence. Although our combined simulations are limited to magnetic-driven turbulence, our results are consistent with the TEMZ simulation, which represents hydrodynamic turbulence. In the following, we list main conclusions from our systematic studies.
\begin{itemize}
\item Turbulence generally leads to stochastic flux and polarization changes, whether it is magnetically driven (as simulated by PIC and polarized radiative transfer) or kinetically driven (as simulated by the TEMZ model).
\item Magnetic turbulence can co-accelerate electrons and protons into very similar power-law distributions.
\item Nonthermal particles accelerated by magnetic reconnection in turbulent plasma are anti-correlated with the local magnetic field strength within magnetic islands.
\item PD and PA from turbulence usually fluctuate around a mean value.
\item Turbulence can lead to smooth or bumpy PA rotations with arbitrary amplitudes and duration, which are not correlated with flares. However, large-amplitude PA swings should be rare due to their stochastic nature.
\item Both flux and polarization are more variable at higher frequencies.
\item The level of variability in light curves and polarization depends on the correlation length of turbulence.
\item The average PD depends on the ratio of the correlation length to the size of the turbulent region.
\end{itemize}

The dependence of the average PD on the turbulence correlation length in the PIC simulations can have interesting physical implications for blazar jets. It is expected that the correlation length is roughly the size of the physical process that drives the turbulence (as also shown in our PIC simulations), such as shocks and both magnetic and hydrodynamic instabilities. Then the average PD can constrain the size of these processes. It is often concluded that turbulence is the main physical mechanism governing fluctuations in flux and polarization during the quiescent state of a blazar. Observations usually find an average quiescent PD of $\sim5$-$ 10\%$ \citep[e.g.,][]{Marscher2021}, corresponding to a physical scale of $\sim 10\%$ of the blazar zone that drives the turbulence. As shown by many magnetohydrodynamic simulations, kink instabilities in the blazar jet mostly concentrate in the central spine, which is only a small part of the jet cross section \citep[e.g.,][]{BarniolDuran2017,Dong2020}. Therefore, the observed low PD may suggest that the turbulence is driven by magnetic instabilities. As shown in our simulations, higher PD corresponds to a larger turbulence correlation length, which may indicate a turbulence driver with a larger physical size. Several works have found that flat-spectrum radio quasars tend to have a higher average PD than BL Lac objects \citep{Angelakis2016,Smith2009,Marscher2021}, which may suggest that they may have a larger kinked central spine if the turbulence is driven by kink instabilities, or a larger internal or recollimation shock if the turbulence is driven by shocks.

While we strive for a comprehensive study, our simulations have a few caveats. First, the turbulence simulation is limited to 2D, which features local magnetic island structures. In 3D, however, the plasma is more disordered \citep{Comisso2021}. Nonetheless, our main results do not necessarily depend on the presence of magnetic islands. We will examine the validity of our results in future 3D simulations. Second, PIC simulations generally extend over a very short physical time scale, which can only track the developing phase of turbulence. Future large-scale simulations that include particle evolution are needed to justify the applicability of these results on typical observational time scales. Finally, our choice of parameters do not have local anisotropy of particles or difference in Lorentz factors, which could inhibit rapid variability in turbulence.

\acknowledgments{We thank the referee for the very constructive and comprehensive review. H.Z. is supported by an appointment to the NASA Postdoctoral Program at Goddard Space Flight Center, administered by Oak Ridge Associated Universities under contract with NASA. The research of A.P.M.\ included in this study was supported by National Science Foundation grant AST-2108622. F.G. gratefully acknowledge the support from Los Alamos National Laboratory (LANL) through its LDRD program, DoE/OFES support, as well as the NASA ATP program. D.G. acknowledges support from the Fermi Cycle 14 Guest Investigator Program 80NSSC21K1938, and the NSF AST-2107802 and AST-2107806 grants. Simulations are carried out on Texas Advanced Computing Center (TACC) Frontera cluster.}

\bibliography{TurbPol1}
\bibliographystyle{aasjournal}



\end{document}